\documentclass[12pt]{iopart} 
\usepackage[top=3cm, bottom =3cm, left = 2.5cm, right=2.5cm]{geometry}
\usepackage{fancyhdr}
\usepackage{graphicx}
\usepackage{color}
\usepackage{xcolor}
\usepackage{epsfig}
\usepackage{float}
\usepackage{multirow}
\expandafter\let\csname equation*\endcsname\relax
\expandafter\let\csname endequation*\endcsname\relax
\usepackage{amsfonts}
\usepackage{amssymb}
\usepackage{amsmath}
\usepackage{tabularx,url,color}
\usepackage{hyperref}
\usepackage{subfigure}
\usepackage{lineno}
\usepackage{upgreek}
\usepackage{comment}
\usepackage{caption}
\usepackage[utf8]{inputenc}
\usepackage[T1]{fontenc}

\begin{document}

\title{The Advanced Virgo Photon Calibrators}

\author{D. Estevez$^{1,2}$, P. Lagabbe$^{2}$, A. Masserot$^{2}$, L. Rolland$^{2}$, M. Seglar-Arroyo$^{2}$, D. Verkindt$^{2}$}
\author{}

\address{$^1$ Institut Pluridisciplinaire Hubert CURIEN, 23 rue du loess - BP28 67037 Strasbourg cedex 2, France}
\address{$^2$ Laboratoire d'Annecy de Physique des Particules (LAPP), Univ.  Grenoble Alpes, Université Savoie Mont Blanc, CNRS/IN2P3, F-74941 Annecy, France}
\ead{dimitri.estevez@iphc.cnrs.fr}

\begin{abstract}
As the sensitivities of LIGO, Virgo and KAGRA detectors improve, calibration of the interferometers output is becoming more and more important and may impact scientific results. For the observing run O3, Virgo used for the first time photon calibrators (PCal) to calibrate the interferometer, using radiation pressure of a modulated auxiliary laser beam impinging on the Advanced Virgo end mirrors. Those optical devices, also used in LIGO, are now the calibration reference for the global gravitational wave detectors network. The intercalibration of LIGO and Virgo PCals, based on the same $\textit{absolute}$ reference called the Gold Standard, has allowed to remove a systematic bias of $3.92\%$ that would have been present in Virgo calibration using the PCal. The uncertainty budget on the PCal-induced displacement of the end mirrors (NE and WE) of Advanced Virgo has been estimated to be $1.36\%$ for O3a and $1.40\%$ on NE PCal (resp. $1.74\%$ on WE PCal) for O3b. This uncertainty is the limiting one for the global calibration of Advanced Virgo. It is expected to be reduced below $\sim 1\%$ for the next observing runs.
\end{abstract}

\maketitle

\markboth{}{}

\section{Introduction}
The burgeoning field of gravitational-wave astronomy has already provided outstanding scientific results in the various domains of astrophysics, cosmology and fundamental physics \cite{GWTC}\cite{H0}\cite{TGR}. The many direct detections of gravitational waves have been performed with the kilometer-scale interferometers of the LIGO-Virgo collaboration which need to deliver accurate detectors' output signals to the scientific community. The calibration of a gravitational-wave detector is a complex procedure which aims at determining the response of the detector to a gravitational wave by mimicking the effect of a gravitational wave in the interferometer. Different quantities are measured throughout the calibration process to provide an accurate reconstruction of the gravitational-wave signal $h(t)$ \cite{O2virgocalib}\cite{O3calibligo}. For instance, the optical response of the interferometer needs to be precisely measured from $\sim 10~$Hz to a few kilohertz, the readout electronics response has to be known at DC and up to a few kHz and the mirrors actuators response which are used in the feedback control loops to keep the interferometer on its working point have to be measured from $\sim 10~$Hz to $\sim 1~$kHz.

The Advanced Virgo detector participated in August 2017 to the observation run O2 and allowed the first triple coincident detection of a binary black holes merger \cite{tripledetect} and the first detection of a binary neutron stars merger with electromagnetic counterpart \cite{binaryneutron}\cite{multimessenger}. During this period, the calibration was using the input laser wavelength as a reference to measure the mirrors electromagnetic actuators response used in the reconstruction of the gravitational wave signal \cite{O2virgocalib}\cite{these_germain}.
For the O3 observation run\footnote{O3 was divided in two periods: O3a from April 1st 2019 to September 30th 2019 and O3b from November 1st 2019 to March 27th 2020.}, from April~1st~2019 to March~27th~2020, the improved sensitivities of the Advanced LIGO and Advanced Virgo detectors required to reduce the calibration uncertainties on $h(t)$. This motivated the use in Virgo of Photon Calibrators as reference to measure the mirrors actuators response \cite{these_estevez}. A calibration tool referred to as \textit{photon calibrator} (PCal) uses photon radiation pressure on suspended interferometer mirrors to induce differential length variations \cite{O3ligopcal}. Moreover, PCals are also used on the Advanced LIGO and KAGRA detectors, hence an intercalibration of the global gravitational wave detectors network based on the same reference could be performed \cite{O3ligopcal}\cite{these_estevez}.  

After an introduction describing the principle of the PCal, this paper gives in section~\ref{sec:expsetup} a detailed description of the PCals that have been implemented on the Advanced Virgo interferometer and describes in section \ref{sec:intercalib} the first work done on PCal intercalibration between Virgo and LIGO. Then, section \ref{sec:uncertainties} presents the uncertainty budget on the end mirror displacement induced by an Advanced Virgo PCal, estimated over the full observing run O3. Eventually, the conclusion in section \ref{sec:conclusion} contains also a discussion about the challenge of gravitational wave detectors calibration for the next observing runs.

\subsection{Photon Calibrator Principle}

The aim of the PCal is to use the radiation pressure of a modulated laser beam, whose modulated output power is well known and controlled, to induce on an end mirror of the interferometer a displacement which translates into a modification of the dark fringe signal at the output port of the interferometer and a measured modification of the reconstructed equivalent gravitational wave strain signal $h(t)$.

The force induced by radiation pressure on the end mirror with a PCal is expressed as:
\begin{equation}
    F_{pcal}(f) =\frac{2\cos(\theta)}{\text{c}}P_{end}(f)
    \label{eq:pcalforce}
\end{equation}
with $\theta$ the angle of incidence of the PCal laser beam impinging on the end mirror, c$~$the speed of light and $P_{end}(f)$ the laser power reflected by the end mirror at the modulated frequency$~f$. This force generates a displacement of the end mirror which is governed by the mechanical response of the suspended optic. This mechanical response is well approximated (within $\pm0.1\%$) in the range 10~Hz to 400~Hz by a simple pendulum transfer function with a resonance frequency at $0.6~$Hz. Indeed, no other mode from the suspensions is expected to couple to the longitudinal displacement above $10~$Hz up to the violin modes around $450~$Hz. The violin modes are not modeled since the control signals sent to the mirrors actuators are notched at these frequencies and have thus a negligible impact on the longitudinal displacement. This means that the end mirror can be considered as a free mass above $10~$Hz with a roll-off of the oscillation proportional to $[m(2\pi f)^2]^{-1}$, with $m$ the mass of the optic. The induced displacement is thus:
\begin{equation}
    x^{free}_{pcal}(f) = -\frac{1}{m(2\pi f)^{2}}\frac{2\cos(\theta)}{\text{c}}P_{end}(f)
    \label{eq:pcalfreemassdisplacement}
\end{equation}

As the PCal system is acting on an end mirror of the interferometer, the laser power noise introduces some unwanted displacement at any frequency, thus an additional noise in the sensitivity of the interferometer. Therefore, a digital system to mitigate the laser power noise of the PCal and to stabilize the modulated output power has been implemented, as described in section \ref{subsec:fastcontrolloop}, so that the remaining broadband noise does not significantly contribute to the Advanced Virgo sensitivity.

\subsection{Mechanical response of the PCal}
In practice, equation \ref{eq:pcalfreemassdisplacement} works well for the Advanced Virgo PCal between $10~$Hz and $400~$Hz. Indeed, contrary to the Advanced LIGO PCal two beams configuration \cite{aligopcal}, the Advanced Virgo PCal uses a single laser beam impinging in the center of the end mirror of the interferometer. Resonant axisymmetric elastic modes of the optic are thus excited and their contribution affects the mechanical transfer function of the PCal above $400~$Hz as it has already been demonstrated in \cite{Hildpcal}.
In Advanced Virgo, the internal deformations of the end mirror are expected to have a significant contribution to the mechanical response in the frequency range of interest from $400~$Hz to $2~$kHz. The first axisymmetric resonance that shows up in the mechanical response is the \textit{drumhead} mode of the end mirror measured around $7813~$Hz.
A  model can be built using the coupling of the \textit{drumhead} mode with the displacement of the end mirror modeled as a second order low-pass filter plus a gain accounting for all the other modes contribution:
\begin{equation}
    H_{tot}(f) = G + \frac{G_d}{1+\frac{j}{Q_d}\frac{f}{f_d}-\Big(\frac{f}{f_d}\Big)^2}
    \label{eq:drumheadcoupling}
\end{equation}
with $G_{d}$ the gain of the \textit{drumhead} mode coupling, $f_d$ the resonant frequency of the mode, $Q_d$ the quality factor and $G$ the frequency independent contribution of the other modes. The effective displacement of the end mirror sensed by the interferometer is thus driven by the following equation:
\begin{equation}
    x_{pcal}(f) = \Big[ -\frac{1}{m(2\pi f)^{2}} + H_{tot}(f) \Big]\frac{2\cos(\theta)}{\text{c}}P_{end}(f)
    \label{eq:pcaldisplacement}
\end{equation}

\begin{figure}[!h]
	\center
	\captionsetup{justification=justified}
	\includegraphics[trim={0cm 1.5cm 0 1cm},clip,scale=0.76]{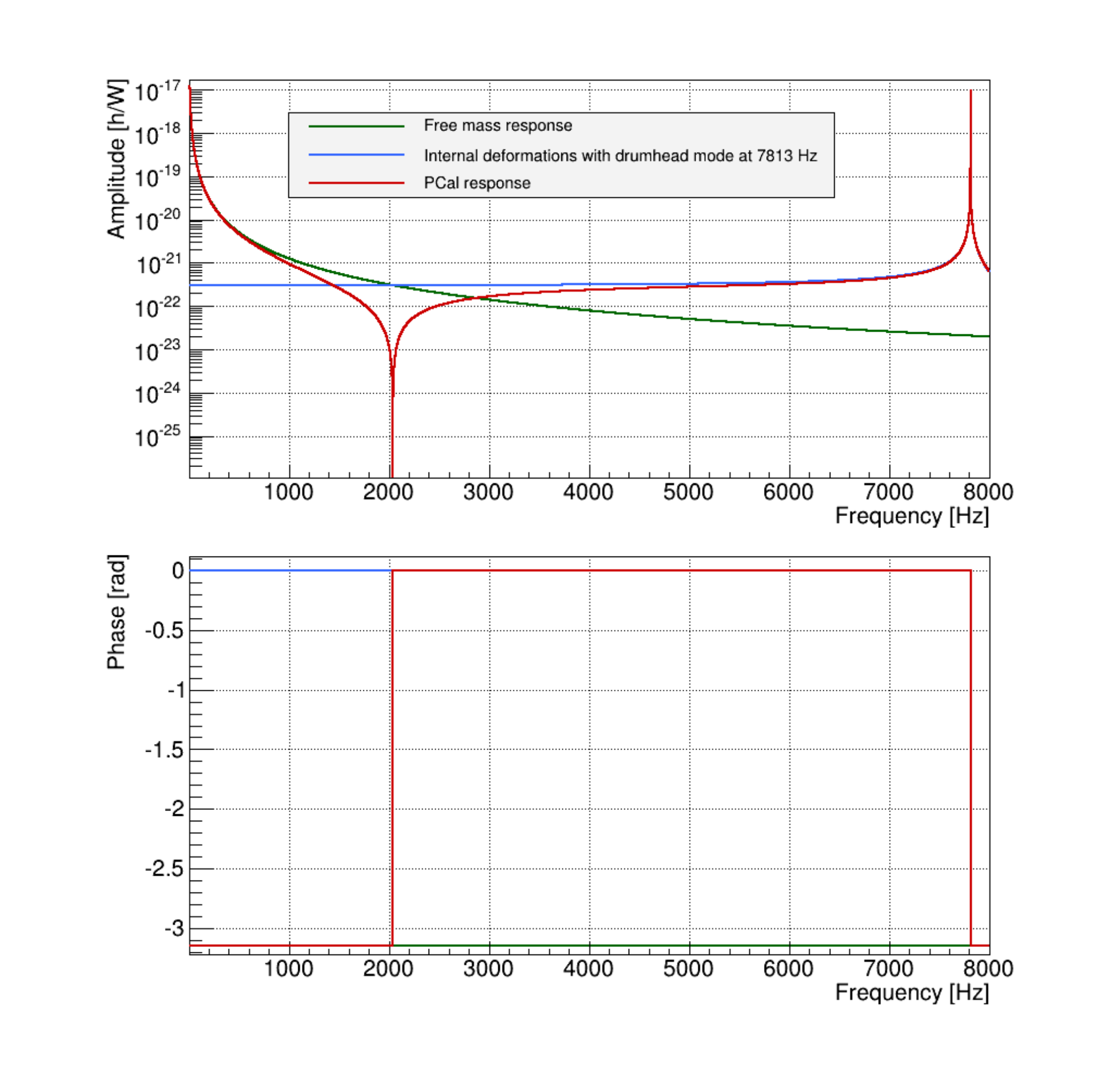} 
	\caption{Expected amplitude and phase of the mechanical response of the Advanced Virgo PCal. This PCal response is the complex sum of the free mass response and of the internal deformations of the end mirror contributions with the first \textit{drumhead} mode excitation of the end mirror at $7813~$Hz.}
	\label{fig:MechanicalPCal}
\end{figure}

For convenience we also define the PCal induced strain as:
\begin{equation}
    h_{pcal}(f) = \frac{x_{pcal}(f)}{L_0}
    \label{eq:pcalstrain}
\end{equation}
where $L_0 = 3000~$m is the nominal length of one arm of the Advanced Virgo interferometer. Figure \ref{fig:MechanicalPCal} shows the expected transfer function from the reflected PCal laser power $P$ to the induced end mirror strain $h_{pcal}$. As the contributions of the free mass response and of the high order modes coupling response are in phase opposition between $10~$Hz and $7813~$Hz, a notch is present at $\sim 2050~$Hz in the PCal response where the amplitude of the two contributions are equal. This means that the interferometer will not be able to sense any displacement of the end mirror induced by the PCal at this frequency. For frequencies above the notch, the PCal response is enhanced by the high order modes coupling instead of falling as $\propto f^{-2}$ as it is the case for a free mass.

\section{Experimental setup}
\label{sec:expsetup}
Two photon calibrators have been installed at the West End (WE) and the North End (NE) stations of the Advanced Virgo interferometer. In addition to being used as reference for the detector's calibration, they allow the verification of the reconstruction of the gravitational wave signal as discussed in section \ref{sec:conclusion}.

\begin{figure}[!h]
	\center
	\captionsetup{justification=justified}
	\includegraphics[trim={0cm 1cm 0cm 1cm},clip,scale=0.7]{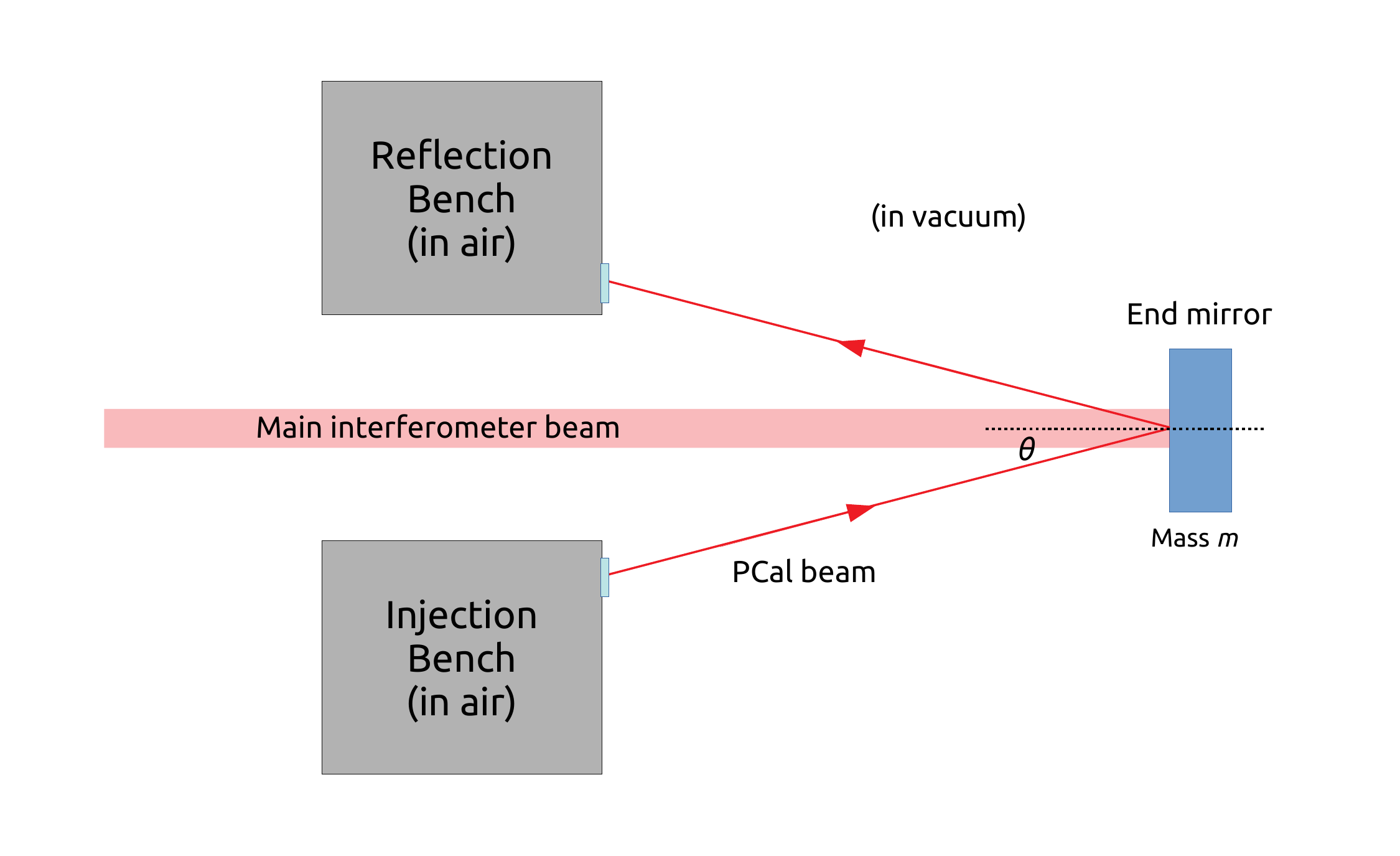} 
	\caption{Schematic of an Advanced Virgo photon calibrator viewed from the top.}
	\label{fig:AdVPCalsetup}
\end{figure}

As shown in figure \ref{fig:AdVPCalsetup}, each PCal setup is composed of two optical benches. The \textit{injection bench} is used to send the laser beam, stabilized in power by a fast digital control loop, to the inner cavity surface of the end mirror. The \textit{reflection bench} is used to measure the power reflected by the end mirror with a Si photodetector sensitive over $1~$cm$^2$. The laser beam hits the center of the end mirror with an angle of incidence $\theta$ of $18.5^{\circ}$.

\subsection{Injection and Reflection Benches}
\label{subsec:pcalbenches}
\label{subsec:timing}
\begin{figure}[!h]
	\center
	\captionsetup{justification=justified}
	\includegraphics[trim={1.8cm 1cm 0cm 1cm},clip,scale=1]{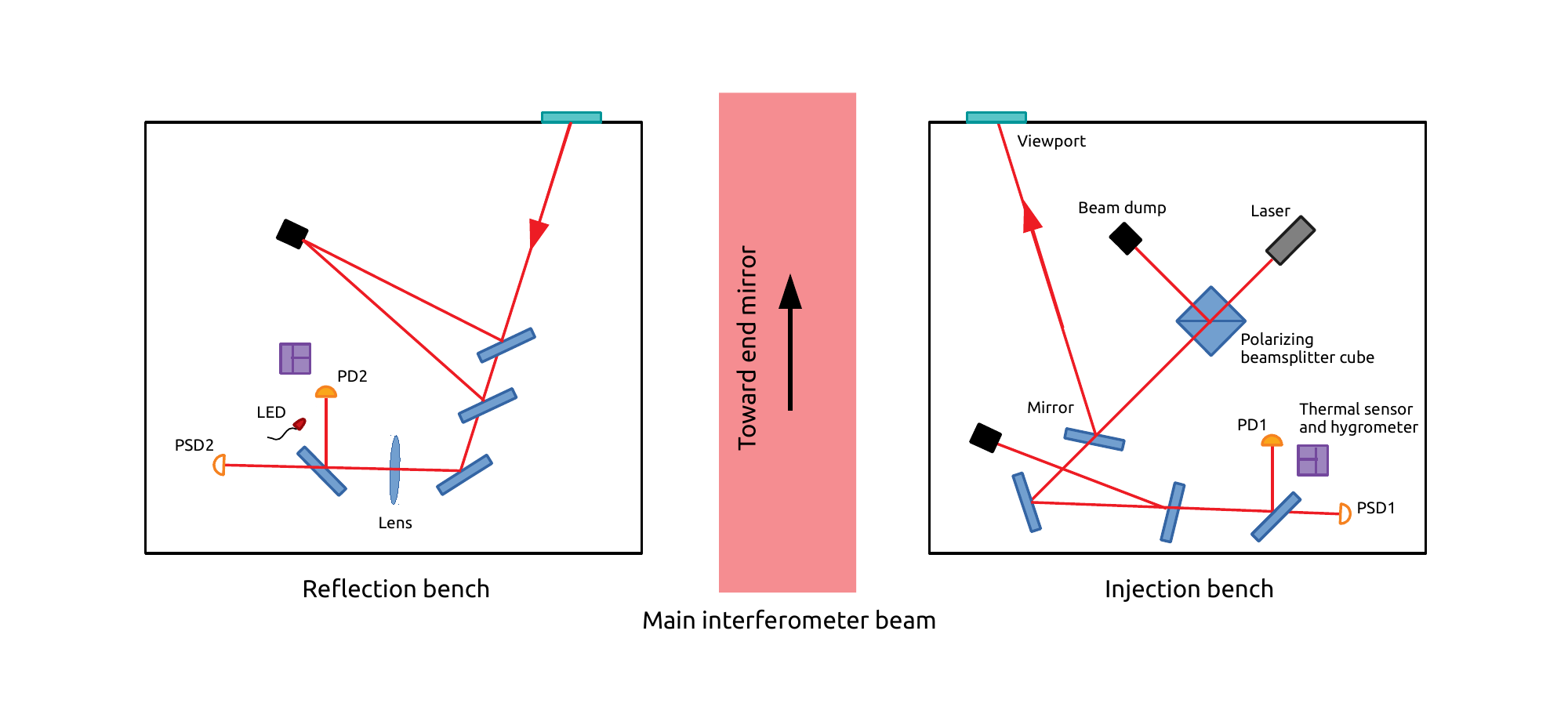} 
	\caption{Detailed schematic of an Avanced Virgo photon calibrator viewed from the top.  The size of the benches is $40~$cm$~\times~40~$cm.
	}
	\label{fig:injreflbench}
\end{figure}

As shown in the optical sketch of the injection and reflection benches of figure \ref{fig:injreflbench}, the Advanced Virgo PCal laser beam is generated with a diode laser source at $1047~$nm. This is the same wavelength as the one used for the Advanced LIGO PCals, which makes easier the laser power intercalibration of both detectors. The PCal laser wavelength is close enough to the main interferometer beam wavelength at $1064~$nm which ensures a high reflectivity of the end mirror of the interferometer (see section \ref{subsec:laserpowercali} for more details) but it is also different enough so that PCal scattered light does not introduce additional noise in the interferometer.

The range of deliverable power goes from $0$ to $3~$W hence the PCal operates at $2~$W to be able to modulate the power up to $\pm1~$W and avoid the non linearity near 0. The modulation of the laser power is directly integrated into the laser driver and can be remotely driven through a front panel BNC connector. The beam is brought to the injection bench through an optical fiber and mounted onto a collimator. It is mainly s-polarized but the first optical component on the injection bench is a polarizing beamsplitter cube that transmits the s-polarized beam and reflects the residual p-polarization down to an optical dump. Most of the laser beam ($\geq 1.9~$W) is then directly sent to the end mirror of the interferometer through a viewport (see section \ref{subsec:laserpowercali} for more details) and only a fraction of the beam ($\sim 5~$mW) is monitored by a photodiode named PD1 after a series of mirrors in order to reduce the amount of light reaching the sensor. This photodiode is the \textit{in-loop} sensor used for the Fast Digital Control Loop described in section \ref{subsec:fastcontrolloop} and its calibration is detailed in section \ref{subsec:laserpowercali}. It is also possible to monitor the laser beam position with a position sensitive detector named PSD1 but it was not used during O3.

The calibration of the photodetectors can be affected by environmental variations of temperature and humidity. Therefore, a thermal sensor and a hygrometer\footnote{The thermal sensor has been implemented at the beginning of O3a and the hygrometer at the beginning of O3b.} have been implemented on the injection bench close to PD1 for monitoring.

The reflection bench houses detectors similar to the ones
of the injection bench. The PCal laser beam reflected by the end mirror of the interferometer reaches the reflection bench through a viewport similar to the one of the injection bench. Only a small fraction of this beam ($\sim 5~$mW) reaches the photodiode PD2 (and another position sensitive detector called PSD2 which has not been used during O3), whose calibration is described in section \ref{subsec:laserpowercali}. Prior to PD2, a lens focuses the laser beam that diverged all along the PCal optical layout. For the same reasons as for the injection bench, a thermal sensor and an hygrometer have been mounted close to PD2.

Eventually, the timing of the PCal system can be measured using a remotely controlled LED flashing a GPS synchronized \textit{1 Pulse Per Second} (1PPS) signal onto PD2. Indeed, the photodiode signal is digitally processed and sampled at $20~$kHz which induces a delay in the readout. Measuring the delay of the 1PPS signal in the $20~$kHz channel allows to calibrate the timing of the PCal, thus the timing of the PCal-induced end mirror motion of the interferometer.

\subsection{Fast Digital Control Loop}
\label{subsec:fastcontrolloop}

The Advanced Virgo PCal operates with an input power of $2~$W. In this state, inherent fluctuations of power in the laser beam occur at every frequency, inducing a broadband displacement of the end mirror. This laser power noise, converted into strain noise, limits the sensitivity of the Advanced Virgo interferometer. Therefore, it has been mitigated to limit its contribution below $10\%$ of the O3 sensitivity.

\begin{figure}[!h]
	\center
	\captionsetup{justification=justified}
	\includegraphics[trim={0cm 1cm 0cm 1cm},clip,scale=0.58]{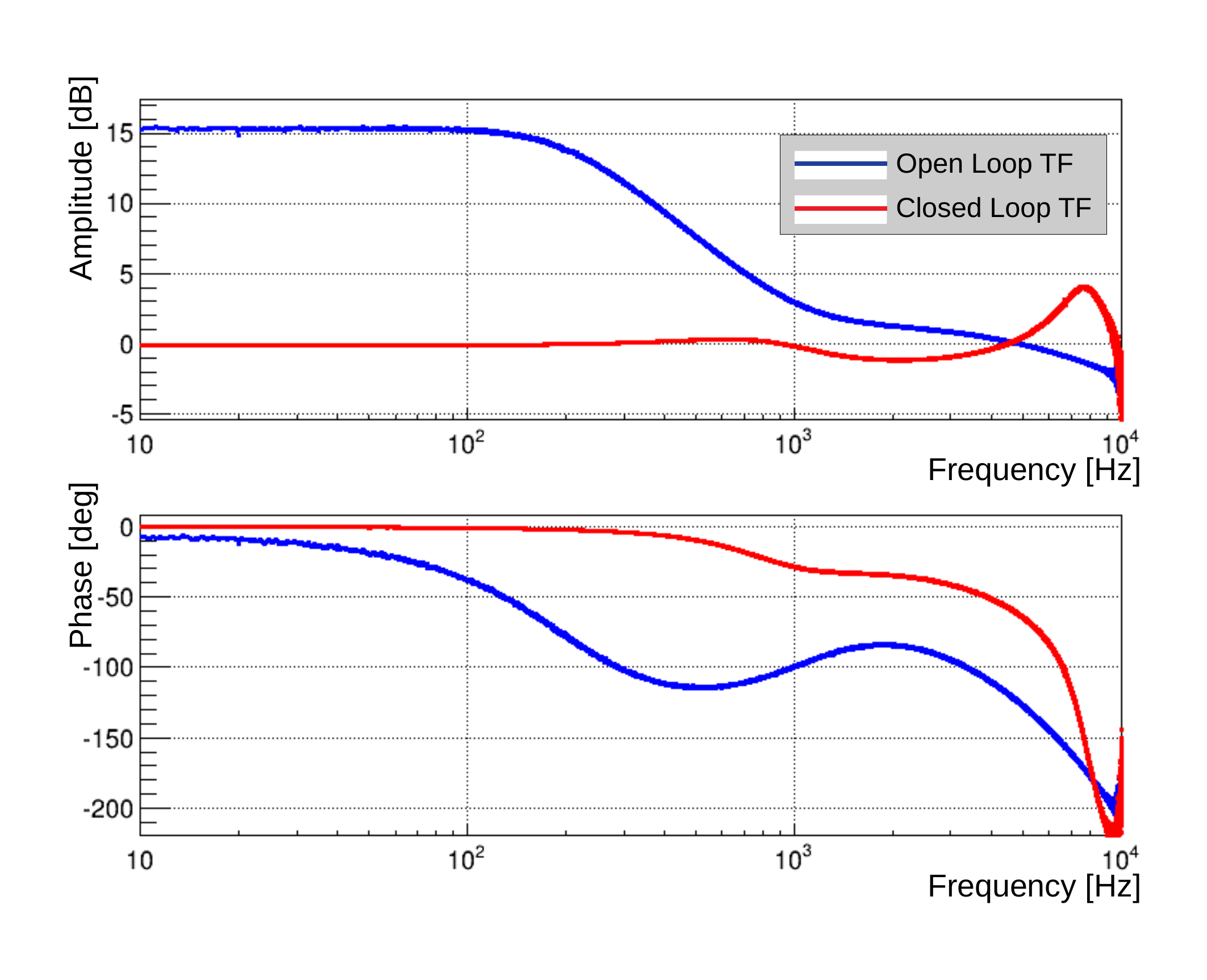} 
	\caption{Open-loop (blue) and closed-loop (red) transfer functions of the PCal Fast Digital Control Loop. The measured unity gain frequency is around $4.8~$kHz and the phase margin is $57^{\circ}$.}
	\label{fig:openclosedloop}
\end{figure}

A fast digital control loop has been implemented to satisfy the above requirement by stabilizing the laser output power at $2~$W. This loop is handled by a real-time process running at $200~$kHz and using an in-loop output power signal witnessed by the photodiode located on the injection bench. The open-loop and closed-loop transfer functions that characterize the fast control loop have been measured (see figure \ref{fig:openclosedloop}). The unity gain frequency is close to $4.8~$kHz with a phase margin of $57^{\circ}$ which ensures a robust control of the system. At $1~$kHz the discrepancy between the requested signal and the output signal is $-0.2~$dB ($-5\%$) and the associated delay is $\tau = 81~\upmu$s ($phase=-29^{\circ}$).

\begin{figure}[!h]
	\center
	\captionsetup{justification=justified}
	\includegraphics[trim={0.5cm 1cm 1cm 1.5cm},clip,scale=0.72]{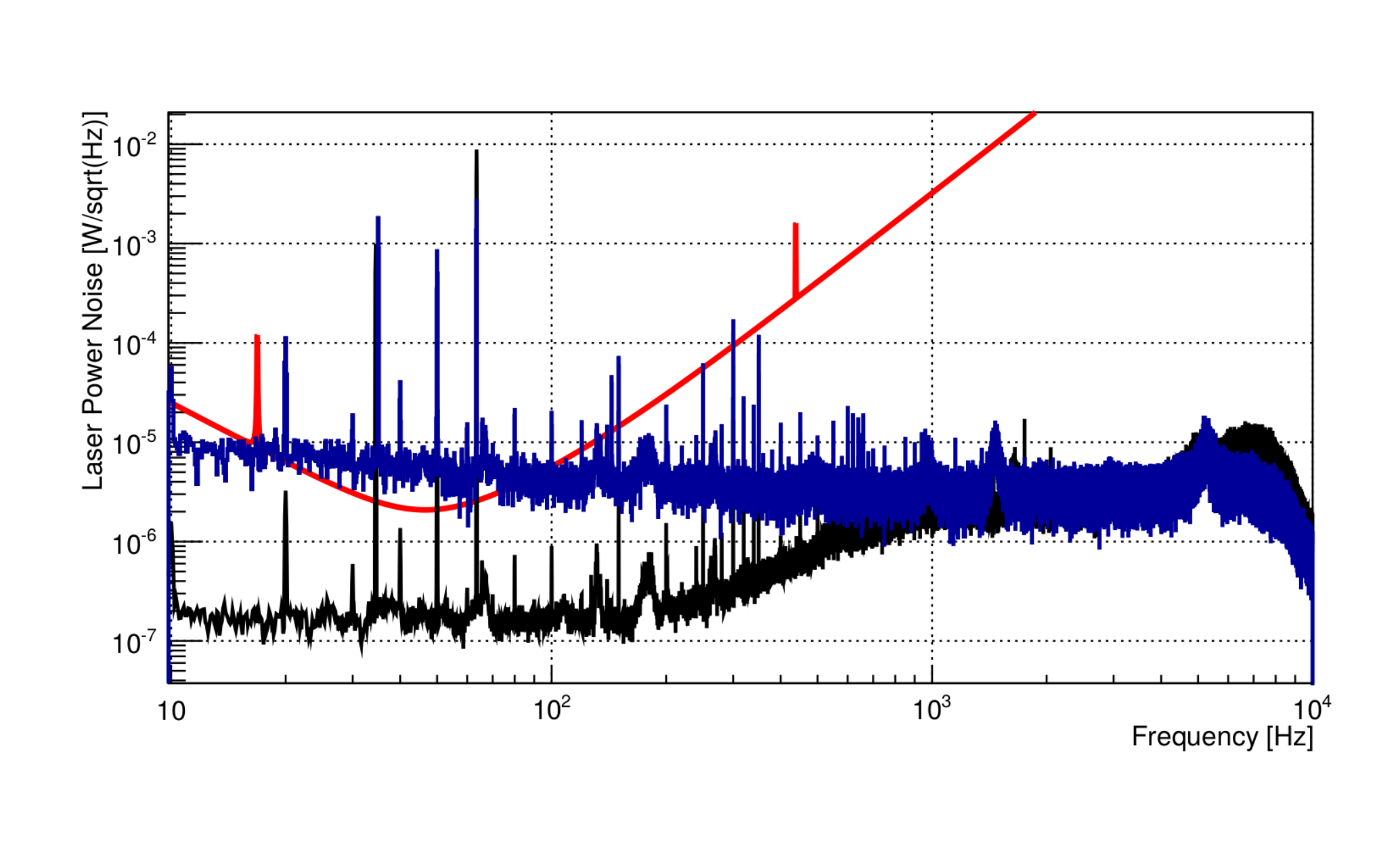} 
	\caption{Laser power noise of the Advanced Virgo PCal without control loop (blue) and with control loop (black). The requirement for the laser power noise of the PCal contribution at $10\%$ of the sensitivity for the observing run O3 is shown in red.}
	\label{fig:LPN}
\end{figure}

The control loop has been running permanently during O3 to stabilize the output power of the laser and to mitigate the laser power noise. Figure \ref{fig:LPN} shows the PCal laser power noise with and without the control loop. Thanks to the loop, the laser power noise during the O3 run was more than one order of magnitude below the requirement set to a maximum contribution at $10\%$ of the sensitivity. It is also worth mentioning that there are three spectral lines remaining above the requirements. One of them is the $50~$Hz signal coming from the distribution of the mains. The two other lines are permanent sinewave excitations at $36.5~$Hz and $60.5~$Hz sent to an end mirror of the interferometer by modulating the laser power of the PCal. They are used to monitor the calibration of the interferometer and the good reconstruction of the gravitational wave signal. Hence, they need to be clearly visible in the Advanced Virgo sensitivity.

\subsection{PCal photodiodes}
\label{subsec:pcalphoto}
The estimation of the laser power reflected by the end mirror is done using the photodiodes on the injection and reflection benches. To do so, those photodetectors have to be carefully calibrated so that the voltage delivered by the sensor receiving the beam can be translated into effective laser power unit reflected by the end mirror. The method for this calibration procedure is to compare laser power measurements with an integrating sphere, so-called Virgo Integrating Sphere (VIS), on the injection and reflection benches recording the laser beam going into the vacuum tower and the laser beam going out of it. In the same time, the photodiode on the injection bench (PD1) collects a fraction of the light of the laser beam and delivers a certain voltage which is recorded and which can then be converted into power unit using the averaged results of VIS measurements from both benches. Once PD1 has been calibrated, VIS is no longer used to calibrate PD2, and the calibration of PD2 is performed comparing the calibrated laser power measured by PD1 against the voltage delivered by PD2. The transfer function between PD1 and PD2 at fixed frequencies between $10~$Hz and $2~$kHz has been monitored once a week during O3. The amplitude was stable within better than $0.02\%$ and the timing of the photodiodes was the same within better than $0.2~\upmu$s. The contribution of these uncertainties are negligible in the overall PCal uncertainty budget.

Before performing the photodiodes calibration, one needs to be careful that the response of the PDs is flat as a function of frequency. Using a LED sending sinusoidal signals and white noise on the photodiodes from $\sim 1~$mHz to $\sim 2~$kHz, an upper limit on the systematic deviation from a flat response has been estimated to $0.04\%$ limited by statistical uncertainties (the limitation comes from the coherence between the LED input signals and the PD output signals). Another aspect to be taken into account is to be sure that VIS power calibration is \textit{absolute}. In section \ref{sec:intercalib}, the \textit{absolute} calibration of VIS is described and this sphere has been used during O3 as the calibration reference for the PCal sensors. 

\section{Intercalibration with LIGO}
\label{sec:intercalib}
One of the main challenge to estimate the displacement of an end mirror of Advanced Virgo induced by a PCal is to determine the PCal laser power reflected by the end mirror. The accuracy on this laser power is the limiting factor of the calibration of the interferometer and impacts the precision of the reconstructed gravitational wave strain provided to data analysis. A pick-off of the reflected laser beam is sensed by photodiode PD2 on the reflection bench and has to be calibrated in an \textit{absolute} manner so that the voltage delivered by the sensor can be precisely converted into absolute power reflected by the end mirror. The calibration of the photodiode is done using the Virgo Integrating Sphere (VIS) which is a Newport $3.3~$in. diameter integrating sphere mounted with a $3~$mm diameter InGaAs photodetector. The linearity of this detector has been measured to $\pm0.4\%$ in the range 0.2~W to 3~W. The method to derive the conversion factor of the photodiode from voltage to power is to simultaneously record the PCal laser power with the integrating sphere and the output voltage of PD1, as described in section \ref{subsec:pcalphoto}. PD2 is then calibrated against PD1 after having removed the integrating sphere of the bench and using the laser beam hitting simultaneously both photodiodes. This procedure requires an \textit{absolute} calibration of the Virgo integrating sphere by carefully chosing an \textit{absolute} calibration reference. We have chosen to use the same reference as LIGO, the so-called LIGO Gold Standard (GS) \cite{O3ligopcal} which is an integrated-sphere powermeter calibrated at the $0.32\%$ level by the National Institute of Standards and Technology (NIST) in Boulder, CO. \cite{GSnist}.

\begin{figure}[h!] 
 \begin{center}
 	\captionsetup{justification=justified}
    \subfigure[]{
	   \includegraphics[trim={1cm 0 1cm 0cm},clip,scale=0.4]{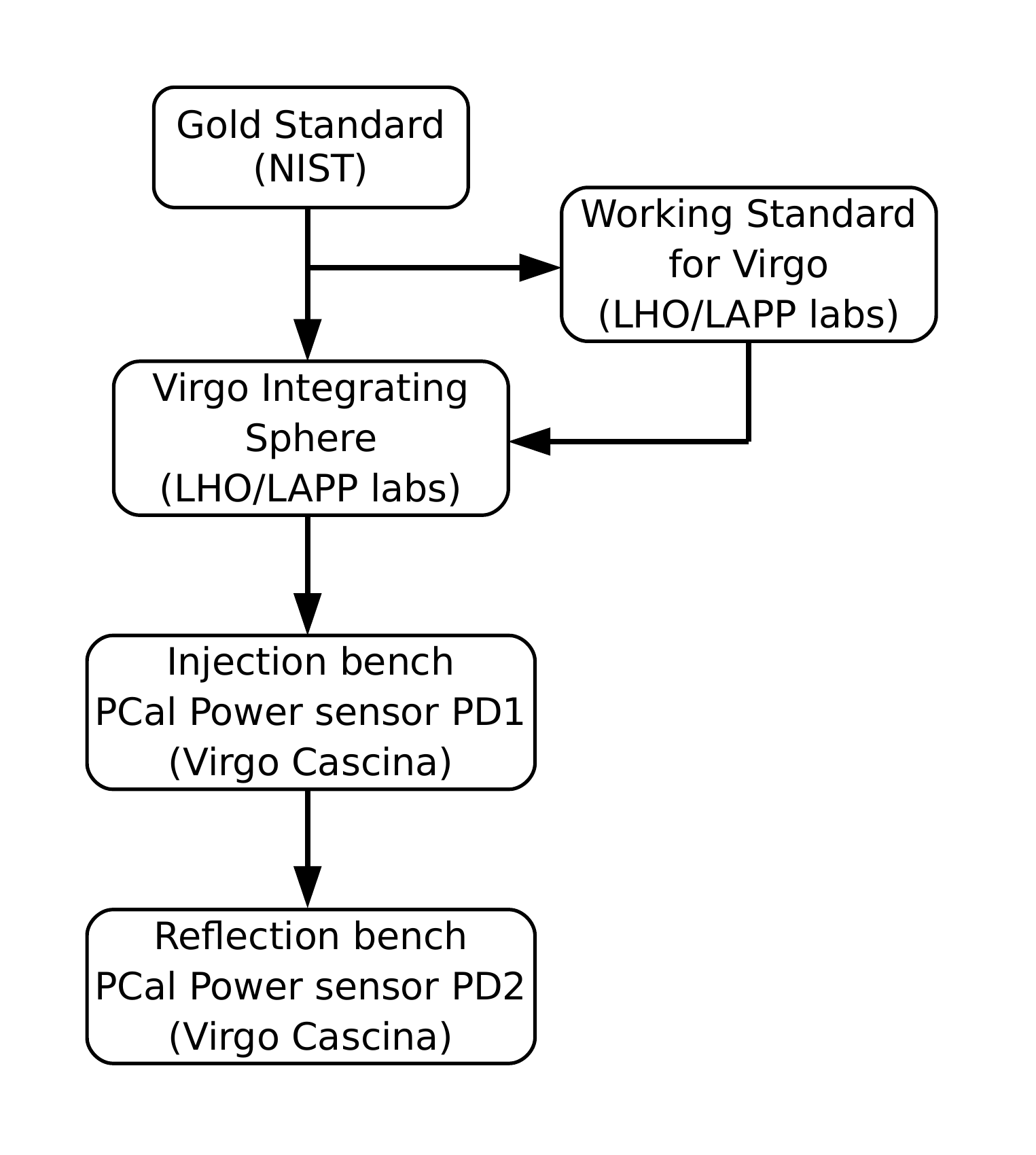} 
	   \label{fig:caltransferligo} }
    \subfigure[]{
       \includegraphics[trim={1cm 0 1cm 0cm},clip,scale = 0.4]{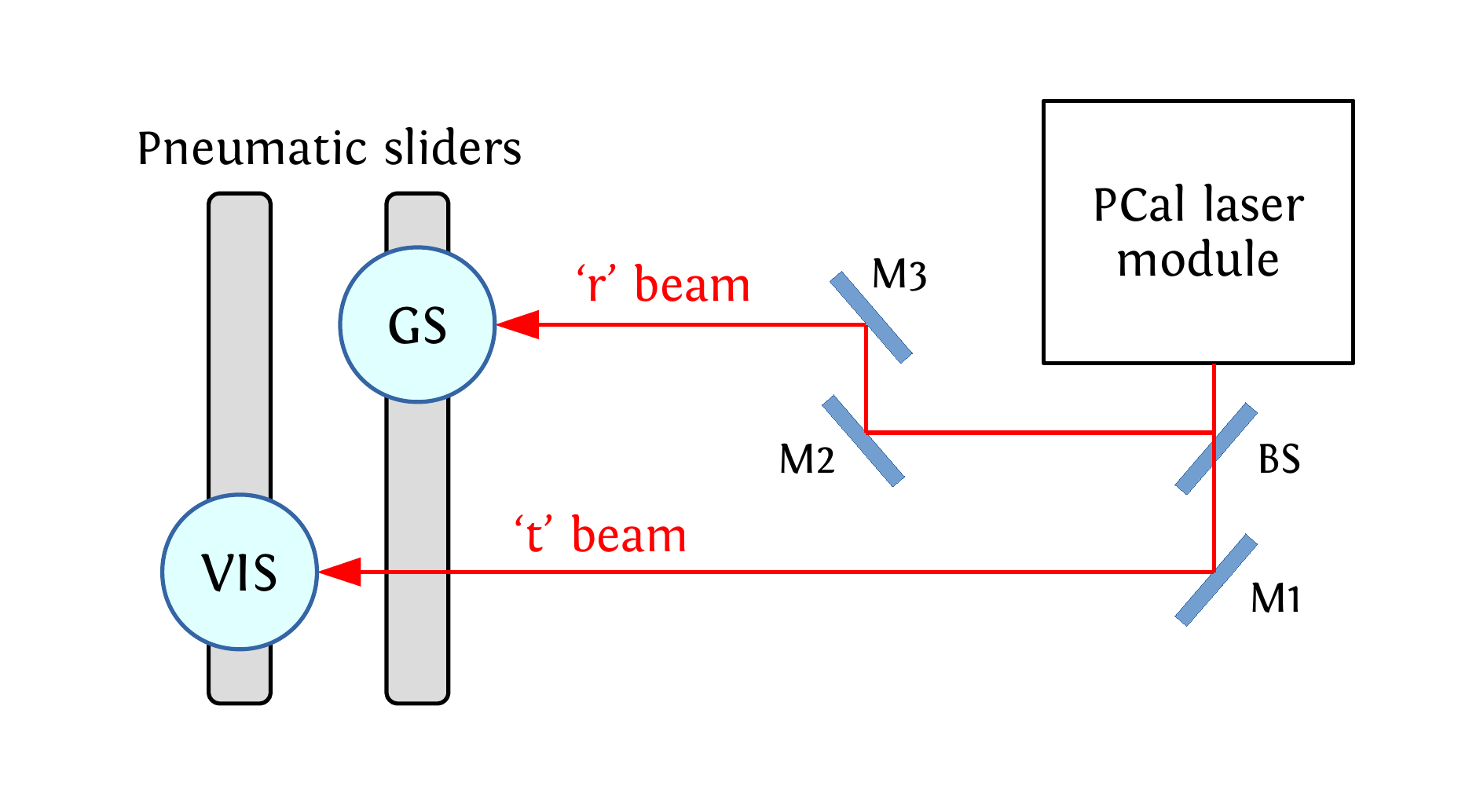}
       \label{fig:caltransfersetup}  }
	\caption{(a) Diagram of the calibration chain made for the laser power calibration of the Advanced Virgo PCal. The LIGO Gold Standard calibrated by NIST stands at the top of the chain and serves as \textit{absolute} reference for the Advanced Virgo PCal power sensors at the bottom of the chain. The location given in brackets indicate where the calibration measurements were performed. (b) Schematic of the optical setup used to calibrate the Virgo Integrating Sphere against the Gold Standard. A PCal laser module is used to generate a laser beam which is then split into two beams with a beamsplitter (BS). The Gold Standard and the Virgo Integrating Sphere are mounted on pneumatic sliders to swap between the reflected (r) and transmitted (t) beams to determine the ratio of the responsivities. This setup is also used to calibrate the Working Standards of Advanced LIGO and KAGRA.}
	\label{fig:intercalisetup}
	\end{center}
\end{figure}

Since LIGO and Virgo are performing a coincident analysis of the calibrated gravitational wave data stream provided by each interferometer, one has to be sure that the relative calibration between each detector does not introduce any bias in the analysis or that at least the putative bias is as small as possible. During the observing run O2, the Free Swinging Michelson technique was used as a reference for Virgo calibration and could not be directly compared to LIGO calibration based on the PCal \cite{aligopcal}. Indeed, the \textit{absolute} reference for LIGO was the Gold Standard and the one for Virgo was the wavelength of the main interferometer laser beam. The decision to use the PCal on Virgo for the observing run O3 was then motivated by the different upgrades performed on the setup which allowed to be confident on its calibration, stability and precision. The use of the PCal was also motivated by the possibility to intercalibrate the PCal laser power between LIGO and Virgo with the Gold Standard. Figure \ref{fig:intercalisetup} shows the calibration chain from GS to the PCal power sensors located on the injection and reflection benches of the Advanced Virgo interferometer and also the optical setup to calibrate VIS against GS.
\subsection{Calibration of the Virgo Integrating Sphere}
The calibration of VIS against GS consists in measuring:
\begin{equation}
\Gamma_{VIS/GS} = \frac{\rho_{VIS}}{\rho_{GS}}
\end{equation}
which is the ratio of the integrating spheres responsivities $\rho$.
The calibration factor $\Gamma_{VIS/GS}$ has been computed as follows:\begin{equation}
\Gamma_{VIS/GS} = \sqrt{\frac{(P_{VIS,r}-P_{VIS,r}^{BG}) \cdot (P_{VIS,t}-P_{VIS,t}^{BG})}{(P_{GS,r}-P_{GS,r}^{BG}) \cdot (P_{GS,t}-P_{GS,t}^{BG})}}
\label{eq:computecalibfact}
\end{equation}
with $r$ and $t$ standing for the reflected and transmitted beam respectively which denotes the position of the spheres on the pneumatic sliders from figure \ref{fig:caltransfersetup}. $P$ stands for the measured powers with incoming laser beam and $BG$ indicates the background measurements with the laser turned off that are subtracted to the measured laser power values. Doing ratios of power measured by GS and VIS eliminates simultaneous laser power variations and swapping their position eliminates the effect of beamsplitter imperfections. As a result this procedure gives access to the ratio of the integrating spheres responsivities $\rho$.

\begin{figure}[!h]
	\center
	\captionsetup{justification=justified}
	\includegraphics[trim={2.2cm 0.5cm 0cm 1.5cm},clip,scale=0.54]{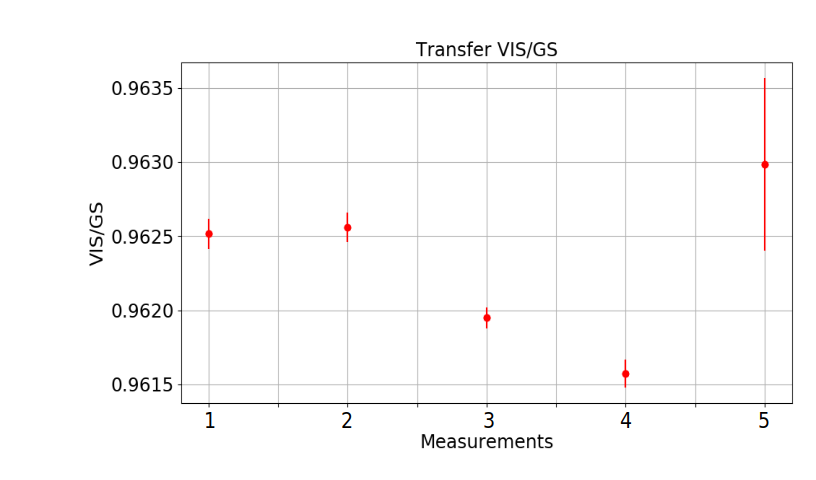} 
	\caption{Calibration factors between the Gold Standard and the Virgo Integrating Sphere. The five points have been measured by averaging 30 sets of 100 s measurements at a fixed laser power $P\sim 0.4$~W on five different days at LHO in February 2019. Only statistical errors are shown. The fifth point has larger error bars suspected to be the consequence of unexpected fluctuations of the laser power servo control.}
	\label{fig:transfervisgs}
\end{figure}

The calibration factor has been computed for five series of measurements performed at LHO in February $2019$ on five different days and the results are shown in figure \ref{fig:transfervisgs}. The average of those measurements is $\overline{\Gamma}_{VIS/GS}~=~0.9623~\pm~0.1\%$ taking into account the dispersion of the points as a systematic uncertainty. The uncertainty on the linearity of the readout of VIS ($\pm~0.4\%$) and the uncertainty on the \textit{absolute} calibration factor measured by NIST for GS ($\pm~0.32\%$) are then added quadratically to the previous uncertainty to give a final estimation of the responsivities ratio $\overline{\Gamma}_{VIS/GS}~=~0.9623~\pm~0.52\%$. This indicates that the power measured by VIS as based on previous calibration was off by almost $4\%$ compared to the power read by GS. For the O3 run, the laser power of the Advanced Virgo photon calibrators were thus calibrated with VIS corrected by $\overline{\Gamma}_{VIS/GS}$.
\subsection{Calibration of the Working Standard for Virgo}
\begin{figure}[!h]
	\center
	\captionsetup{justification=justified}
	\includegraphics[trim={2cm 0.5cm 0cm 1.2cm},clip,scale=0.45]{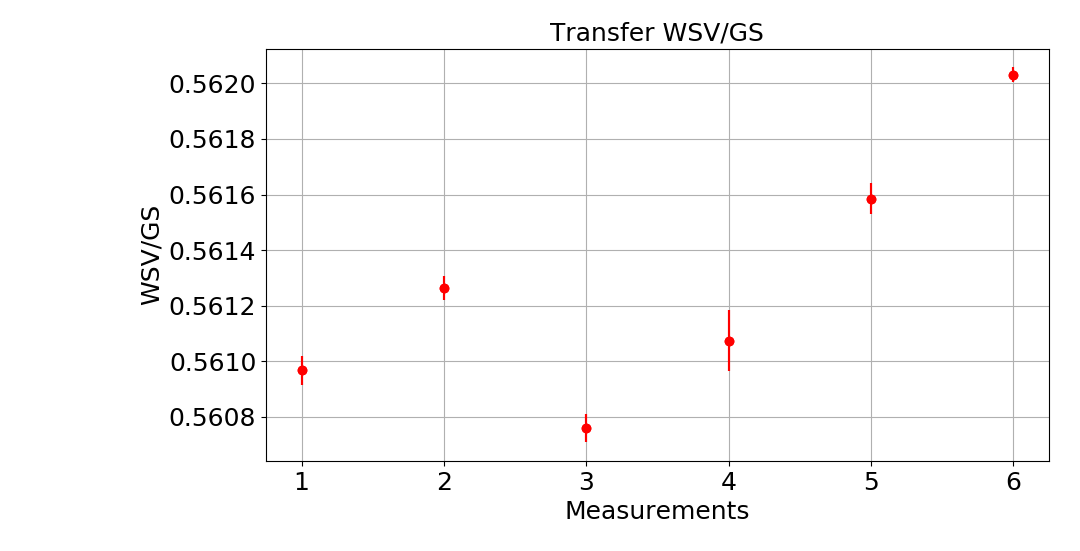} 
	\caption{Calibration factors between the Gold Standard and the Working Standard for Virgo. The six points have been measured by averaging 30 sets of 100 s measurements at $P\sim 0.4$~W on six different days at LHO in February 2019. Only statistical errors are shown.}
	\label{fig:transferwsvgs}
\end{figure}

Since VIS is used on Virgo to calibrate the PCal it cannot be compared against GS (which stays at LHO) as often as needed to check the stability of the calibration factor. A Working Standard for Virgo (WSV) similar to the Working Standards used on LIGO and KAGRA has thus been mounted at LHO with the aim of staying at LAPP\footnote{Laboratoire d'Annecy de Physique des Particules}, where the Virgo PCal is developed, so that VIS could be compared against WSV during O3. This integrating sphere WSV was calibrated against GS with the same setup used to calibrate VIS. Figure~\ref{fig:transferwsvgs} shows the calibration factor corresponding to the ratio of the responsivities of WSV and GS computed as in equation \ref{eq:computecalibfact} for six series of measurements performed at LHO in February $2019$ on six different days. The averaged calibration factor is  $\overline{\Gamma}_{WSV/GS}~=~0.5613~\pm~0.34\%$ with systematic uncertainties. For technical reasons, the ratio is not close to $1$ since the photodiode transimpedance of WSV has been divided by a factor $2$ compared to the one of GS in order to meet the requirements of Advanced Virgo PCal laser power.
\subsection{Intercalibration stability}
\begin{figure}[!h]
	\center
	\captionsetup{justification=justified}
	\includegraphics[trim={0cm 0cm 0cm 0.8cm},clip,scale=0.32]{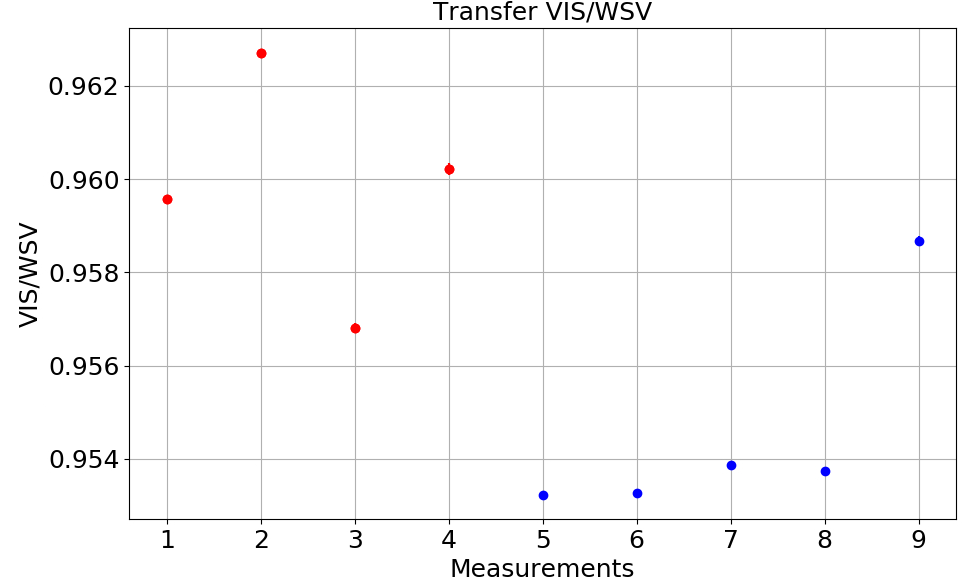} 
	\caption{Calibration factors between the Working Standard for Virgo and the Virgo Integrating Sphere. The red points have been measured by averaging 30 sets of 100 s measurements at $P\sim 0.4$~W on four different days at LHO in February 2019. The blue points have been measured by averaging 1 set of 3600 s measurements at a fixed laser power $P\sim 1$~W on five different days at LAPP (four days in June 2019 and one day in October 2019). Only statistical errors are shown.}
	\label{fig:transferviswsv}
\end{figure}

The ratio of the responsivities of VIS and WSV has also been measured at LHO in the same experimental conditions as for the previous ratios described in the above sections. The value of this ratio is the reference from which the stability of VIS calibration is estimated and has to be monitored at LAPP during O3. Therefore, a similar optical setup to calibrate the integrating spheres as the one at LHO has been mounted at LAPP during O3 to check the stability of VIS calibration. Since the acquisition tools are different from LHO to LAPP we used a voltage calibrator to calibrate our voltage readout at LAPP at the level of $0.007\%$. Figure \ref{fig:transferviswsv} shows the four measurements of $\Gamma_{VIS/WSV}$ performed at LHO in February 2019 and the five measurements performed at LAPP in June and October 2019. The variation of the points around the mean value is $\pm0.5\%$ which is thus the value used during O3 to characterize the stability in time of the intercalibration between LIGO and Virgo.

Investigations to understand the systematic uncertainties related to these measurements are needed for the next observation run O4 in order to find solutions to improve the stability in time of the intercalibration. More measurements of Virgo spheres against the Gold Standard will also have to be performed to strengthen the confidence in the calibration factors.

\section{Uncertainties}
\label{sec:uncertainties}
The total uncertainty on the end mirror displacement induced by the PCal arises from the determination of all the parameters from equation \ref{eq:pcaldisplacement}. The main contribution to this total uncertainty comes from the estimation of the laser power reflected by the end mirror. Indeed, many factors have to be taken into account from the \textit{absolute} calibration of VIS with GS down to the Advanced Virgo PCal sensors calibration.

\subsection{Calibration of the PCal photodiodes}
\label{subsec:laserpowercali}
The power measurements on the injection and reflection benches with VIS to calibrate the photodiodes are done outside the vacuum tower containing the end mirror. The laser beam is thus affected by the optical losses of the viewports between the tower and the PCal benches. Hence, to precisely estimate the power reflected by the ETM, the viewport losses have to be characterized.

The viewports are tilted by $6^{\circ}$ downwards with respect to the center of the ETM and they have parallel faces coated on both in-air and in-vacuum surfaces with 1064 nm broadband anti-reflective coatings whose power reflectivity had been estimated to be around $R = 0.05\%$ per coating at $1047~$nm. The absorption of the viewports is negligible compared to their reflectivity so that one can reasonably assume that all the losses come from the reflected light. Moreover, the power reflectivity coefficient of the end mirror at $1047~$nm with an angle of incidence of $18.5^{\circ}$ is greater than $99.99\%$ and is thus not affecting the total optical efficiency of the PCal. Therefore, the power reflected by the end mirror $P_{end}$ can be expressed as:
\begin{equation}
    P_{end} = P_{inj} (1-R)^2 = \frac{P_{ref}}{(1-R)^2}
\end{equation}
where $P_{inj}$ and $P_{ref}$ are the laser power respectively measured on the injection and reflection benches. $P_{end}$ can thus be approximated, at the first order, as the average of $P_{inj}$ and $P_{ref}$. This is what has been used for the photodiodes calibration. We expect the optical efficiency $\eta = P_{ref}/P_{inj}$ to be around $0.998$ considering the viewport losses meaning that the uncertainty on $P_{end}$ should be within $0.2\%$.

However, when the measurements were performed it was found that the measured laser power on the injection bench was varying at maximum by $0.8\%$, depending on the position of VIS between the viewport and the last mirror before the viewport. On the reflection bench, the measured laser power did not depend on the position of VIS between the viewport and the first mirror of the reflection bench. The main difference between these two measurements is the size of the laser beam which is bigger at the reflection bench. We thus expect that if the beam size is smaller than a specific size, a non negligible amount of light entering VIS is reflected back outside the sphere. This effect would give an under estimation of the measured power. We found indeed a position on the injection bench where the power was almost $0.5\%$ smaller than the power on the reflection bench and another position where the power was $0.3\%$ higher. The measurements were thus performed with VIS located where the beam size is the largest on the injection bench, and the laser power was then $0.3\%$ higher than the power measured on the reflection bench which is closer to the value we would expect from the optical efficiency.

Since we could not estimate properly the optical efficiency of the PCal due to the effect mentioned above, we decided to keep a conservative uncertainty of $1\%$ for this measurement resulting from $0.8\%$ due to power variations depending on VIS position plus $0.2\%$ considering the expected optical efficiency. This is the dominant uncertainty in the final error budget of the PCal and it will have to be tackled and characterized more accurately in the future. A possible study would be to measure the laser power with VIS at different positions on an optical bench and with different laser beam sizes.

Table \ref{tab:laserpower} summarizes the sources of systematic uncertainties on the estimation of the PCal laser power reflected by the end mirror of the interferometer, from the top (the Gold Standard) to the bottom (the PCal sensors) of the laser power calibration chain. Then, the total relative uncertainty on $P_{end}$ is computed as:
\begin{equation}
    \frac{\sigma_P}{P_{end}} = \Big[\sum_i \Big(\frac{\sigma_{x_i}}{x_i}\Big)^2 \Big]^{1/2} 
\end{equation}
with $x_i$ the different parameters estimated as sources of systematic uncertainties in the laser power calibration chain. 
\begin{table}[!h]
\centering
\captionsetup{justification=justified}
\begin{tabular}{|c|c|}
\hline 
Parameter & $1\sigma$ Uncertainty \\ 
\hline 
GS responsivity (2018) & $0.32\%$ \\ 
VIS linearity & $0.4\%$ \\
VIS/GS responsivity ratio & $0.1\%$ \\
VIS/WSV responsivity ratio & $0.5\%$ \\
Voltage calibrator & $0.007\%$ \\
VIS position and optical efficiency & $1\%$ \\
PD frequency dependant response & $0.04\%$ \\
\hline
Power reflected by the end mirror & $1.24\%$ \\
\hline
\end{tabular}
\caption{\label{tab:laserpower} Summary of the sources of systematic uncertainty on the estimation of the PCal laser power reflected by the end mirror of the interferometer.}

\end{table}

\subsection{Geometrical parameters}
\label{subsec:geometricalparam}
The geometrical parameters are also contributing to the overall uncertainty on the end mirror displacement. Above the resonant frequency of the end mirror suspension, the displacement of the optics induced by a PCal is inversely proportional to the mass of the optics. In Advanced Virgo, the mass of an end mirror including the ears, anchors and magnets has been estimated to $42.37\pm0.02~$kg from drawings and material density. Unfortunately, the value used in the PCal calibration analysis was $42.3~$kg and the uncertainty on the mass was thus increased to $\pm0.07~$kg which gives a $1\sigma$ uncertainty of $\pm0.17\%$ on the PCal-induced end mirror displacement.

The angle of incidence of the PCal laser beam hitting the end mirror has been evaluated, with optomechanical constraints from the drawings, to be $\theta = 18.5^{\circ}$. This angle is limited by the diameter of the viewports. This diameter is $63~$mm and the beam has been centered on the viewports better than $\pm10~$mm. The 1$\sigma$ uncertainty on the angle of incidence is thus treated as a Type-B uncertainty \cite{typeunc} contributing in the cosine as $\pm0.12\%$.

\begin{figure}[h!] 
 \begin{center}
 	\captionsetup{justification=justified}
    \subfigure[]{
	   \includegraphics[trim={1cm 0cm 1cm 1cm},clip,scale=0.6]{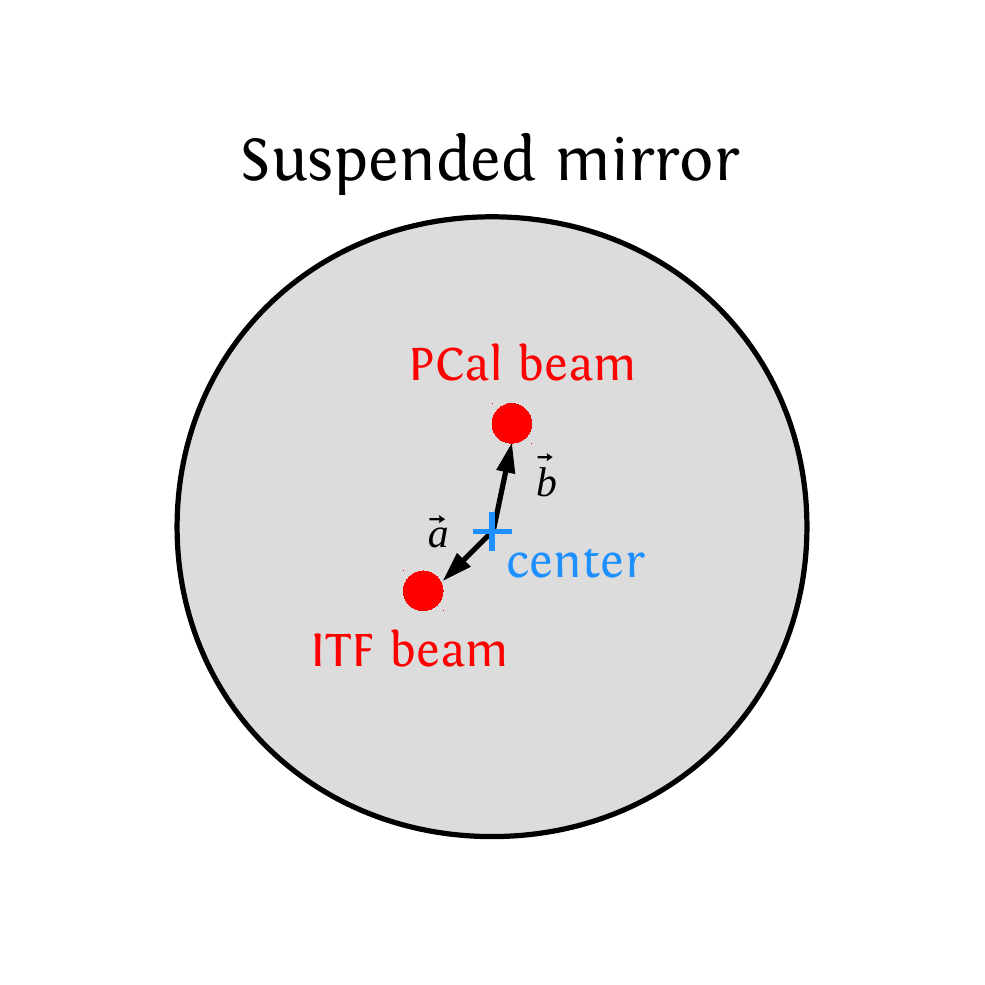}  
	   \label{fig:rotationpcal} }
    \subfigure[]{
       \includegraphics[trim={0cm 0 1cm 1.7cm},clip,scale = 0.52]{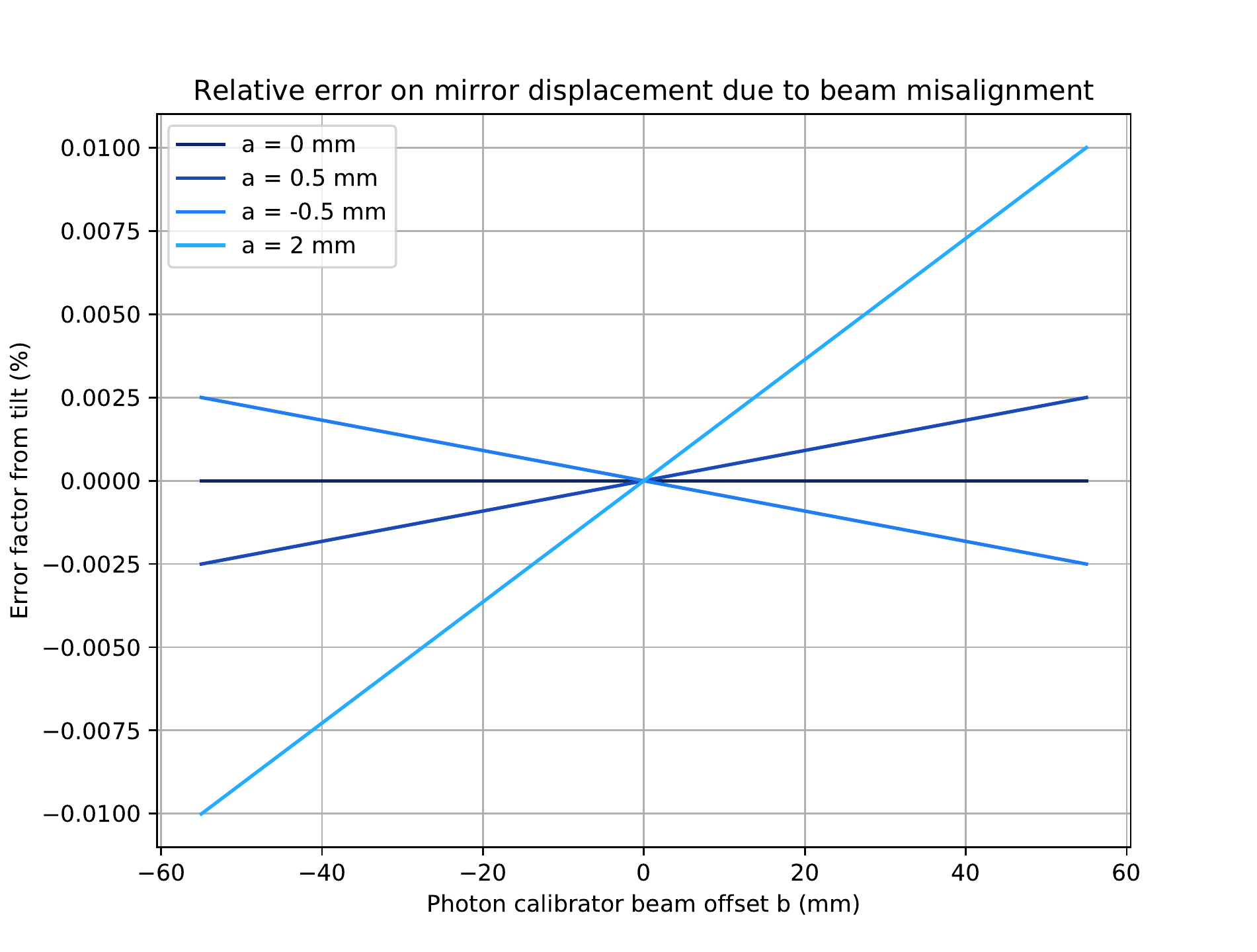}
       \label{fig:relativerotationerror}  }
	\caption{(a) Schematic of a suspended end mirror of the interferometer with the main interferometer (ITF) laser beam and the PCal laser beam spots shifted from the center of the optics by $\vec{a}$ and $\vec{b}$ respectively. The miscentering of the beams is exagerated in this figure for clarity purpose. (b) Relative error on the end mirror displacement due the tilt of the optics induced by a miscentering of the main interferometer beam and the PCal beam. The vectors positioning the beams are assumed to be collinear in order to maximize the error.}
	\label{fig:pcalgeometry}
	\end{center}
\end{figure}

Unwanted rotation of the end mirror can be caused by a torque induced by a miscentering of the PCal laser beam and the main interferometer laser beam as shown in figure \ref{fig:rotationpcal}. This effect changes equation \ref{eq:pcalfreemassdisplacement} by adding an end mirror rotation term of the form $-\frac{\vec{a}\cdot\vec{b}}{I(2\pi f)^2}$ in the mechanical transfer function, with $I$ the rotational moment of inertia of the end mirror and $\vec{a}$ (resp. $\vec{b}$) the vector from the center of the mirror to the position of the main interferometer laser beam spot (resp. PCal laser beam spot) on the optic.

In Advanced Virgo, the centering of the main interferometer beam is controlled to be better than $\pm0.5~$mm and the PCal laser beam is centered better than $\pm20~$mm due to optomechanical constraints. Considering the worst case scenario where the scalar product between the two vectors is extremum (both beam spots are shifted in the same direction), the miscentering of the beams leads to a relative error of $\pm0.001\%$ on the displacement of the end mirror due to the tilt of the optics.

\begin{table}[!h]
\centering
\captionsetup{justification=justified}
\begin{tabular}{|c|c|}
\hline 
Parameter & $1\sigma$ uncertainty \\ 
\hline 
Mass of the end mirror & $0.17\%$ \\ 
Angle of incidence (cosine) & $0.12\%$ \\
Rotation of the optic & $0.001\%$ \\
\hline
Total  & $0.20\%$ \\
\hline
\end{tabular}
\caption{\label{tab:geometricunc} Summary of the sources of systematic uncertainties on the geometrical parameters of the PCal-induced end mirror motion equation.}
\end{table}

\subsection{Mechanical response of the PCal}
\label{subsec:mechanicalresponse}
As mentioned in previous sections, the PCal laser beam hits the center of the end mirror of the interferometer and thus excites axisymmetric internal modes of the optic. The model of the PCal-induced motion of the end mirror is given by equation \ref{eq:pcaldisplacement} where $H_{tot}(f)$ is the last item of the equation that we have to characterize. As we do not expect the contribution of the \textit{drumhead} mode to be dominant in the internal deformations below $2~$kHz and the quality factor being greater than $10^6$, we make the following approximation for $f\ll f_d$:
\begin{equation}
H_{tot}(f) \approx G_{tot}
\end{equation}
where $G_{tot}$ is the frequency independent gain of the internal deformations including the contribution of all the excited high order modes of the optic.\\  
The measurement of this gain has been done by measuring the frequency of the notch in the mechanical response of the PCal around $2~$kHz as shown in the simulated response in figure~\ref{fig:MechanicalPCal}. Indeed, the notch is the result of the free-mass response having the same amplitude as the internal deformation of the mirror but in phase opposition. The idea is thus to compare a PCal-induced strain $h_{pcal}$ on the end mirror, taking into account only the simple pendulum model $H_{p}$, with the reconstructed strain of Advanced Virgo interferometer $h_{rec}$. Taking the transfer function from the PCal strain to the reconstructed strain will reveal the discrepancy between both strains which arise from the unmodeled internal deformations. It is important to notice that this measurement depends on the reconstruction of the gravitational wave strain and may introduce unwanted bias in the measurement. We thus assume that if there is a bias in $h_{rec}$ around the notch frequency band it is a constant bias on the amplitude and it is not frequency dependent so that the shape of the measurements is unchanged.

\begin{figure}[h!] 
 \begin{center}
 	\captionsetup{justification=justified}
    \subfigure[WE]{
	   \includegraphics[trim={0.9cm 1cm 1.4cm 1cm},clip,scale=0.6]{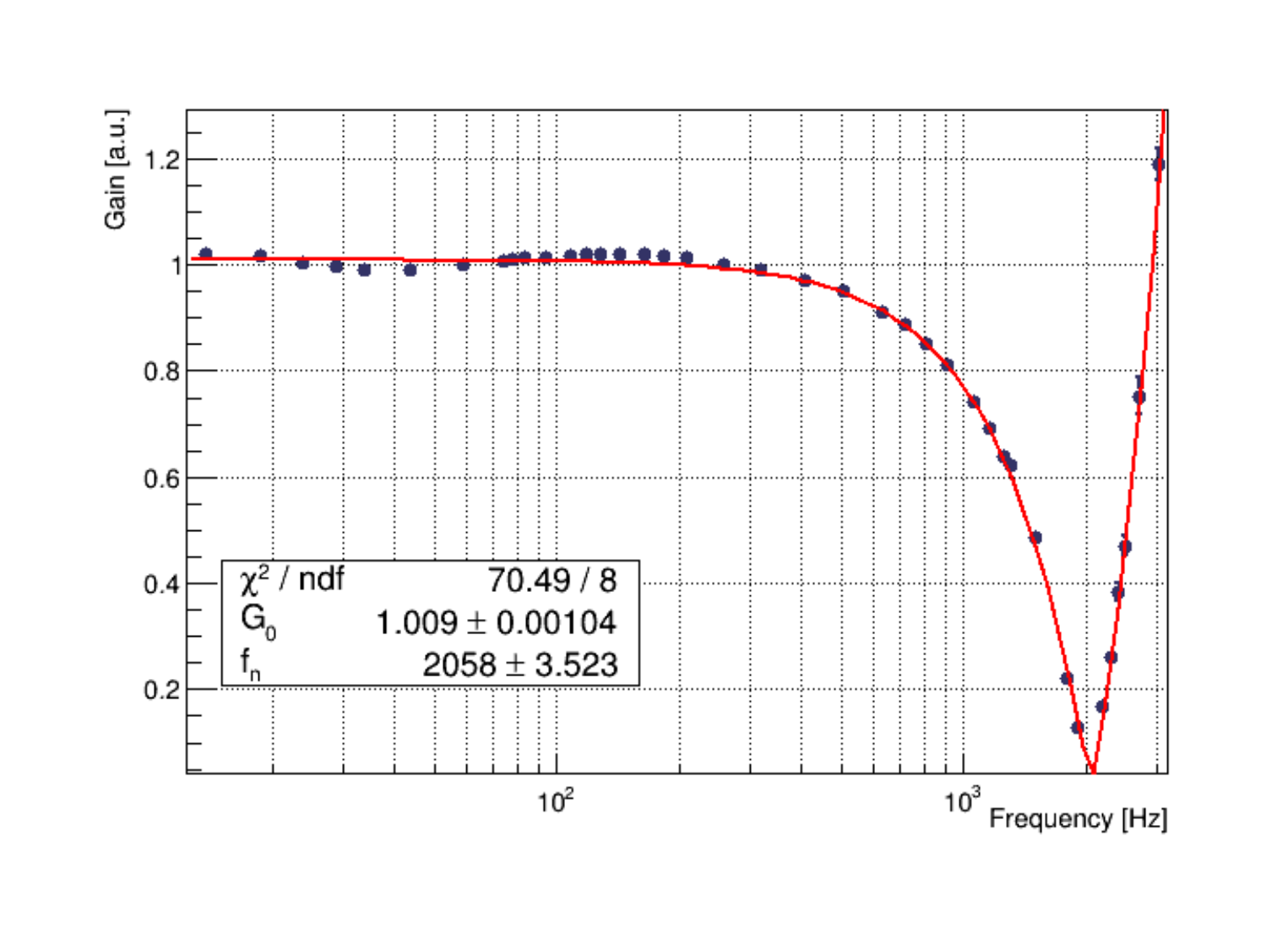}  
	   \label{fig:notchpcalWE} }
    \subfigure[NE]{
       \includegraphics[trim={0.9cm 1cm 1.4cm 1cm},clip,scale = 0.6]{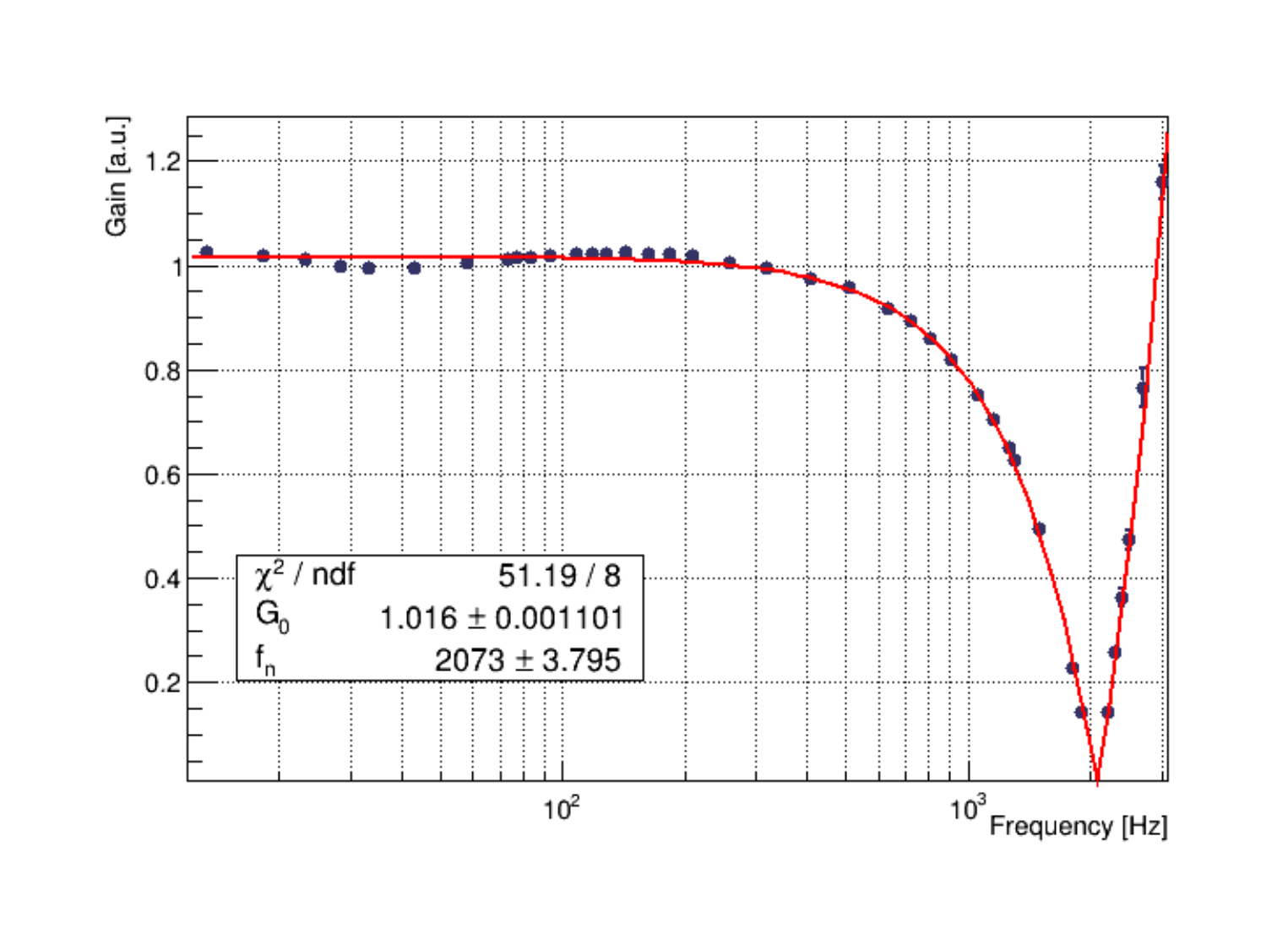}
       \label{fig:notchpcalNE}  }
	\caption{Measured amplitude of $h_{rec}/h_{pcal}$ for WE PCal (a) (NE PCal (b)) and the associated fit.}
	\label{fig:notchpcal}
	\end{center}
\end{figure}

The quantity that we want to measure and fit can be expressed as:
\begin{align}
\frac{h_{rec}}{h_{pcal}} & \propto \frac{H_{p} + H_{tot}}{H_{p}} \nonumber \\
						& \propto G_0\left(1-\frac{f^2}{f_n^2}\right)
\end{align}
where $G_0$ is a global gain and $f_n$ is the notch frequency. They are the two parameters of the fit. The total gain of the internal deformations of the optic is given by:
\begin{equation}
    G_{tot} = \frac{1}{4\pi^2 m f_n^2}
\end{equation}
Figure \ref{fig:notchpcal} shows the amplitude of the transfer function for both WE and NE PCals. The total gain $G_{tot}$ for both PCals have been renormalized in unit of strain per watt for convenience and have been estimated to:
$$G_{tot}^{WE} = (2.98\pm0.01) \times 10^{-22} ~\text{h/W}$$
$$G_{tot}^{NE} = (2.94\pm0.01) \times 10^{-22} ~\text{h/W}$$
\subsection{Stability during O3a and O3b}
The calibration of the PCal photodiodes has been monitored during O3 to look for any other sources of systematic uncertainties and to check the calibration stability in time. The responsivity of the Advanced Virgo PCal photodiodes depends on the temperature as $0.5\%/^{\circ}$C at $1047~$nm. A monitoring of the surrounding temperature during O3 has thus been done to evaluate the impact on the overall PCal uncertainty budget. Since the temperature variations do not follow a gaussian distribution we treated the change in photodiode responsivity as a Type-B uncertainty assuming a rectangular distribution of temperatures over the whole range of variations. The highest range of temperature variations was found to be $0.65^{\circ}$C on NE PCal \textit{reflection bench} which resulted in a Type-B uncertainty of $\pm 0.1\%$ on the photodiode calibration, to be added to the PCal uncertainty budget.

\begin{figure}[h!] 
 \begin{center}
 	\captionsetup{justification=justified}
    \subfigure[WE]{
	   \includegraphics[trim={1.2cm 0.8cm 1.3cm 1cm},clip,scale=0.43]{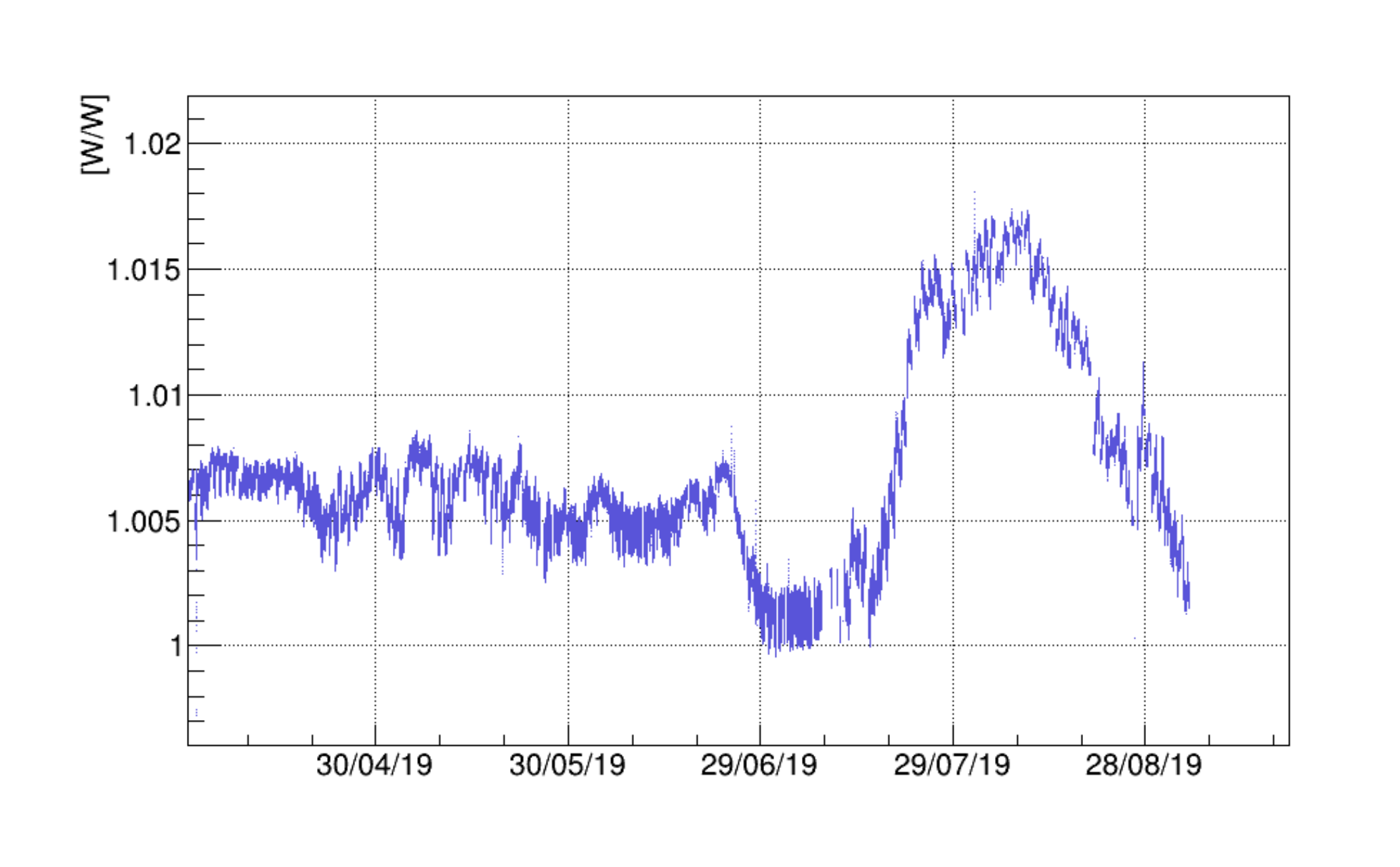}  
	   \label{fig:WEPD1PD2} }
    \subfigure[NE]{
       \includegraphics[trim={1.2cm 0.8cm 1.3cm 1cm},clip,scale = 0.37]{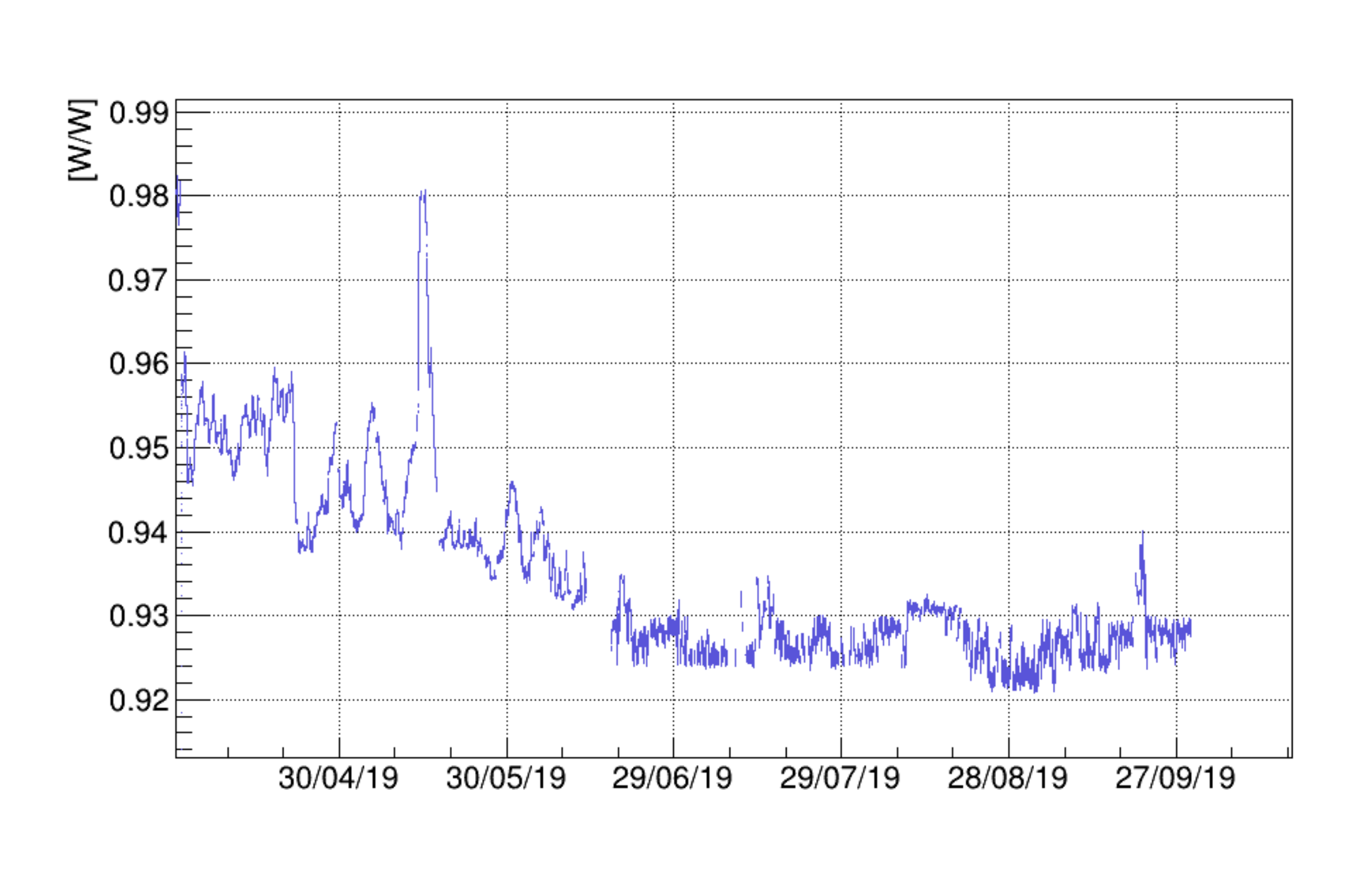}
       \label{fig:NEPD1PD2}  }
	\caption{Ratio of PD1 and PD2 signals over O3a at the frequency of the calibration signals. (a) The calibration signal frequency is $60.5~$Hz on WE and (b) $63.5~$Hz on NE.}
	\label{fig:variationsPD1PD2}
	\end{center}
\end{figure}

Some variations of photodiodes power calibration larger than the expected ones due to temperature variations were seen between PD1 and PD2 during O3a as shown in figure~\ref{fig:variationsPD1PD2}.

\begin{figure}[h!] 
 \begin{center}
 	\captionsetup{justification=justified}
    \subfigure[WE PD1 (left) and WE PD2 (right).]{
	   \includegraphics[trim={1cm 1cm 0cm 1cm},clip,scale=0.86]{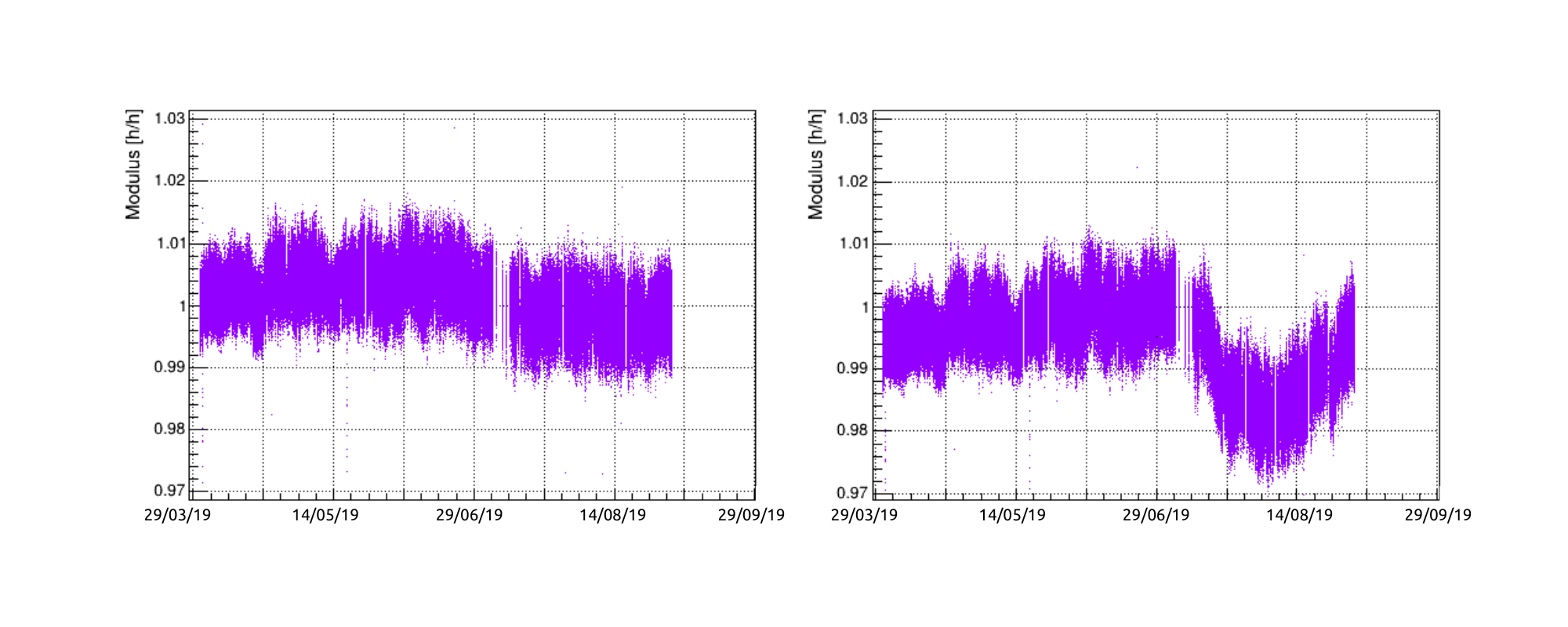}  
	   \label{fig:WEhrechpcal} }
    \subfigure[NE PD1 (left) and NE PD2 (right).]{
       \includegraphics[trim={1cm 1cm 0cm 1cm},clip,scale = 0.86]{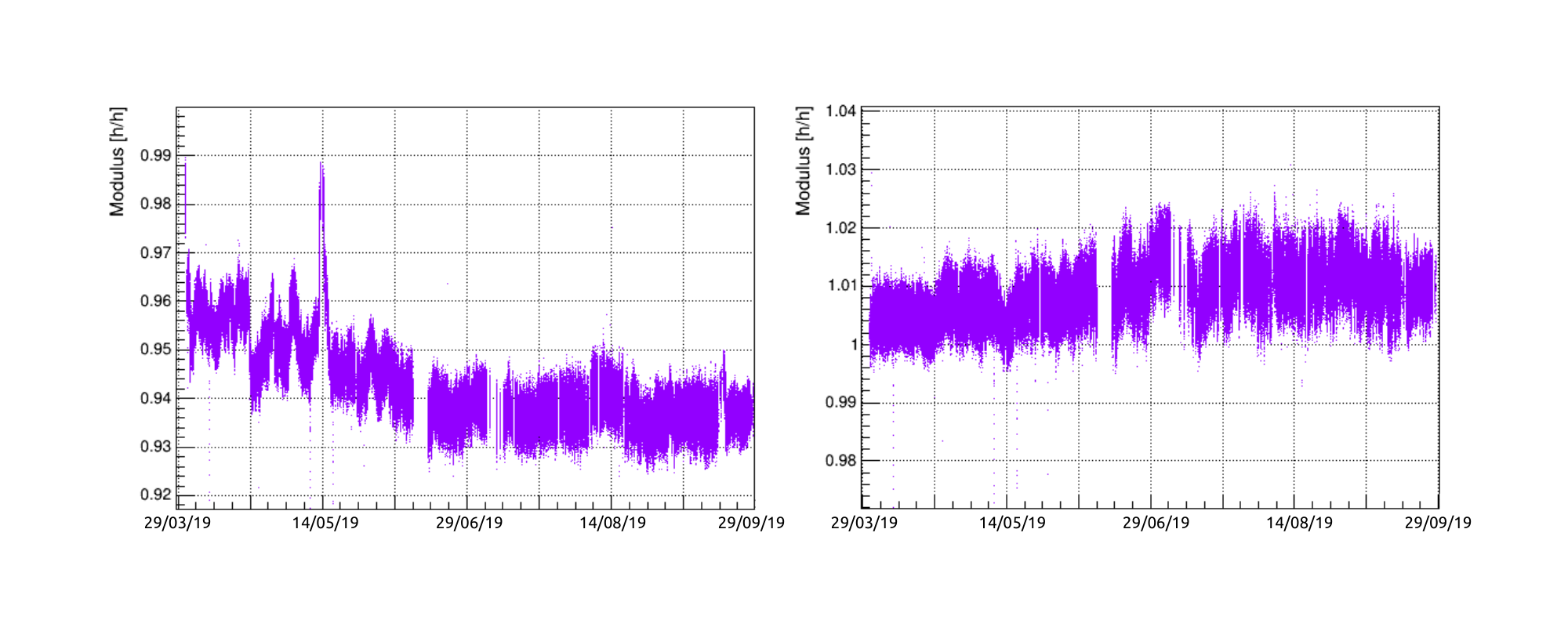}
       \label{fig:NEhrechpcal}  }
	\caption{Ratio of $h_{rec}$ and $h_{pcal}$ for both PD1 and PD2 signals on both PCals at the frequency of the calibration signals during O3a. (a) Calibration signal frequency is $60.5~$Hz for WE and (b) $63.5~$Hz for NE.}
	\label{fig:hrechpcalO3a}
	\end{center}
\end{figure}

\begin{figure}[h!] 
 \begin{center}
 	\captionsetup{justification=justified}
    \subfigure[WE PD1 O3a]{
	   \includegraphics[trim={1.2cm 1cm 1.2cm 1cm},clip,scale=0.45]{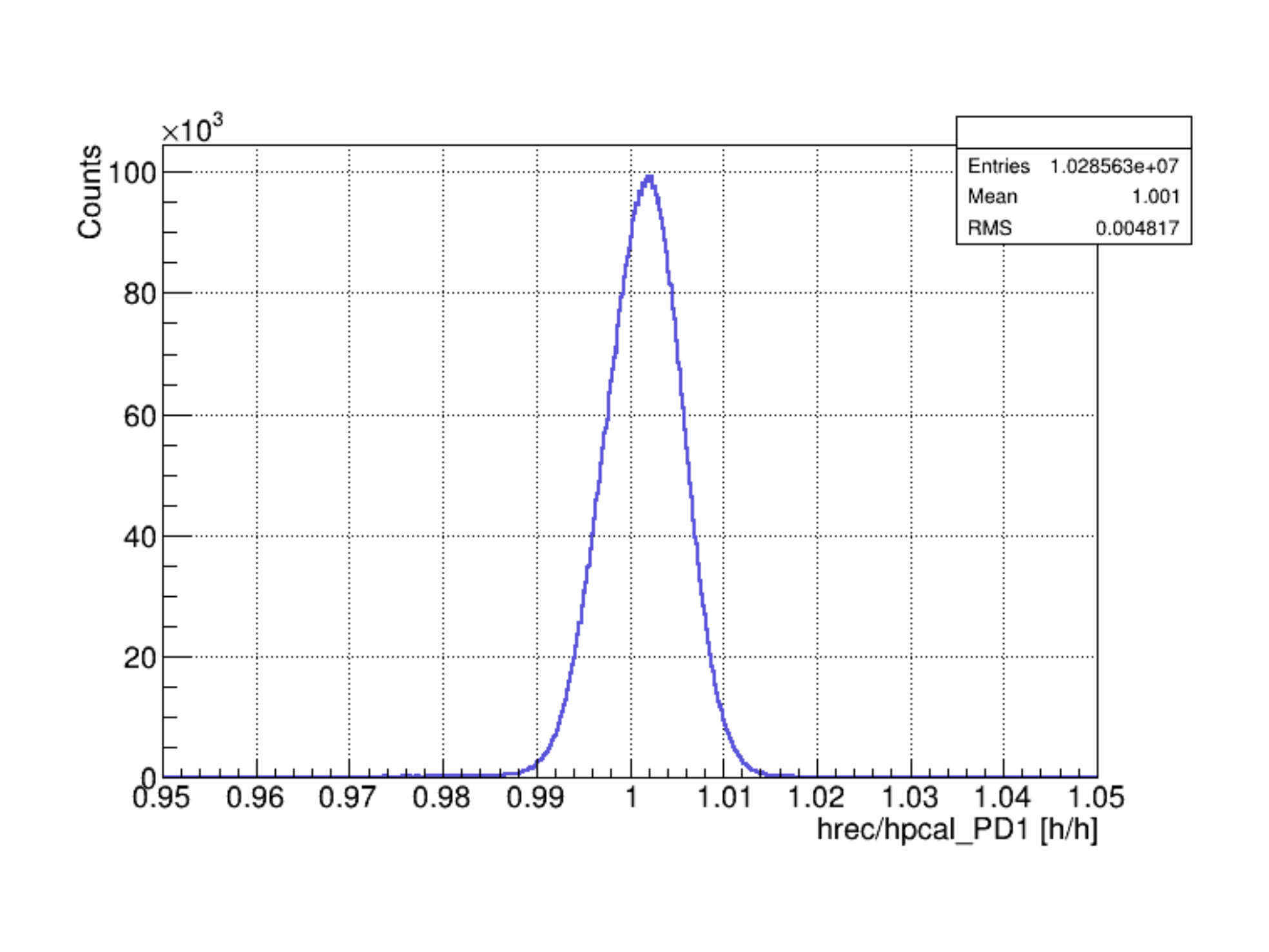} 
	   \label{fig:distribpcalWEO3a} }
    \subfigure[NE PD2 O3a]{
       \includegraphics[trim={1.2cm 1cm 1.2cm 1cm},clip,scale = 0.45]{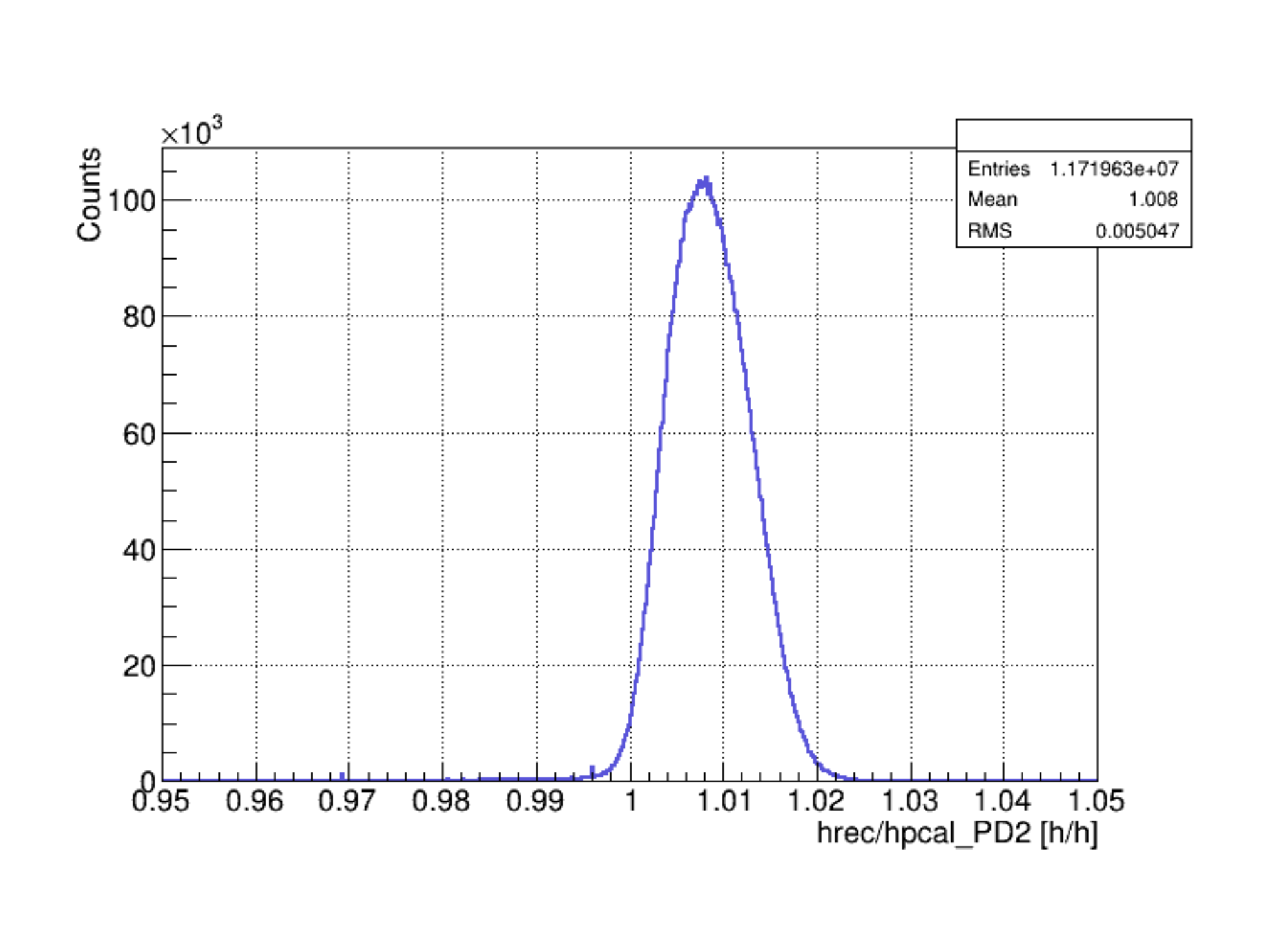}
       \label{fig:distribpcalNEO3a} }
           \subfigure[WE PD1 O3b]{
	   \includegraphics[trim={1cm 1cm 1cm 0cm},clip,scale=0.4]{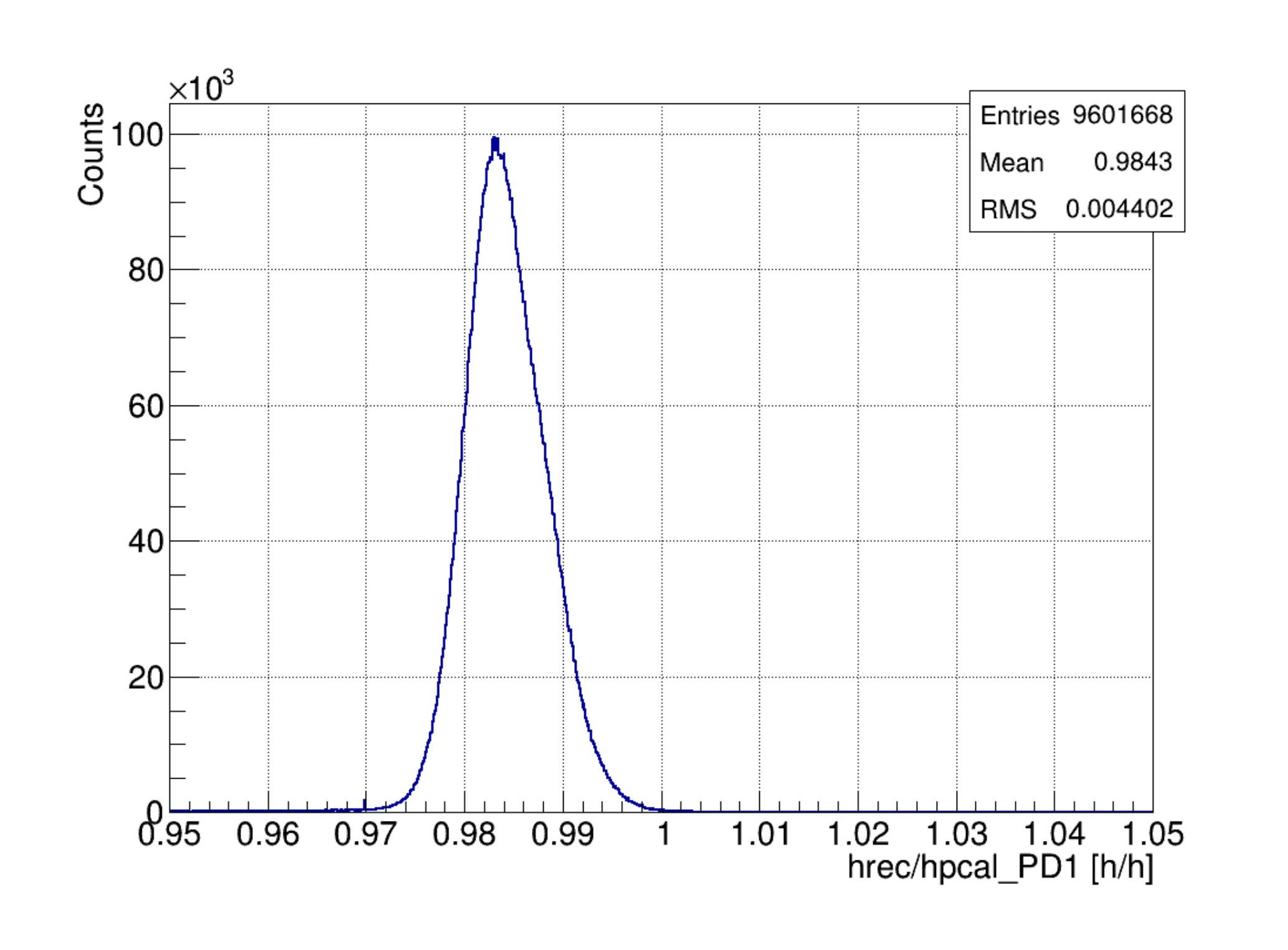} 
	   \label{fig:distribpcalWEO3b} }
    \subfigure[NE PD2 O3b]{
       \includegraphics[trim={0cm 0cm 0cm 0cm},clip,scale = 0.45]{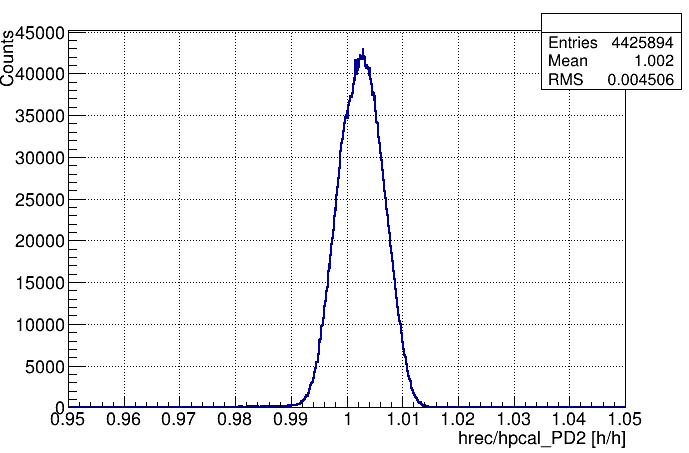}
       \label{fig:distribpcalNEO3b} }
	\caption{Distribution of $h_{rec}/h_{pcal}$ using (a) WE PD1 and (b) NE PD2 during O3a and (c) WE PD2 and (d) NE PD2 during O3b. The standard deviation of the distributions is$~\sim 0.5\%$ and gives an upper limit on the stability of the PCal calibration over O3a and O3b for a given calibration of the photodiodes.}
	\label{fig:distribhrechpcalO3}
	\end{center}
\end{figure}

The monitoring and the analysis of these variations were done using calibration signals of the PCal. Those signals are sinewave excitations sent to the end mirrors of the interferometer by modulating in amplitude the laser beam of the PCal. A long term study was done over O3a comparing the reconstructed signal $h_{rec}$ from the output of the interferometer against the signal $h_{pcal}$ reconstructed from the PCal photodiodes. Figure \ref{fig:hrechpcalO3a} shows the ratio of $h_{rec}$ over $h_{pcal}$ for both photodiodes on both PCals during O3a. The photodiodes PD2 on WE and PD1 on NE are the two photodiodes that contribute the most to the variations seen on figure \ref{fig:variationsPD1PD2}. Only the photodiodes PD1 on WE and PD2 on NE were thus used to estimate the PCal-induced end mirrors motion during O3. The uncertainty on their calibration stability has been assessed using the width of the distributions of $h_{rec}$ over $h_{pcal}$. Figure \ref{fig:distribhrechpcalO3} shows these distributions for O3a and O3b. Those distributions account for calibration variations of $h_{pcal}$ but also the ones of $h_{rec}$, thus only an upper limit on the uncertainty on the $h_{pcal}$ stability can be drawn from this analysis. Therefore the $1\sigma$ uncertainty on the stability of the photodiodes power calibration has been conservatively estimated to $\pm0.5\%$ over the O3a period.

In between O3a and O3b\footnote{One month break in the observation run of LIGO and Virgo from October 1st to November 1st 2019.}, the NE and WE driver lasers had to be repaired after an electrical failure. Only the WE driver laser was mounted back in time for the start of O3b and the NE driver laser has been reinstalled later during O3b in January 2020. A few days before O3b, the WE PCal set-up had to be realigned and WE PD1 recalibrated. The new measured calibration factor for WE PD1 differed by $+1.3\%$ from the one of O3a. This difference exceeds the uncertainty of $0.8\%$ stated in section \ref{subsec:laserpowercali} due to VIS positioning on the optical bench and therefore was significant enough to be corrected for O3b. One can see in figures \ref{fig:distribpcalWEO3a} and \ref{fig:distribpcalWEO3b} that the mean values of the distributions differ by $\sim 1.7\%$ and with similar standard deviations of $0.5\%$. It has been found that this difference between both distributions is due to the recalibration of WE PD1 and to relative humidity changes. We thus chose to extend the uncertainty on the WE PCal PD1 calibration from $0.5\%$ for O3a to $1.2\%$ for O3b as explained further in this section. 

NE driver laser was mounted back on the 21st of January 2020 but not recalibrated to be able to compare the new set-up to the previous one and also with WE PCal. Figures \ref{fig:distribpcalNEO3a} and \ref{fig:distribpcalNEO3b} show that the mean values between both distributions differ by $0.5\%$ with similar standard deviations of $0.5\%$. Both distributions being compatible without performing any recalibration of the photodiodes we added them together to get the uncertainty on the stability of NE PCal for O3b which becomes $0.6\%$.

Table \ref{tab:stabilityunc} summarizes the uncertainty contributions of the temperature variations and \textit{other sources} that affect the PCal laser power stability. Investigations to understand the \textit{other sources} of variations which affect the stability in time of the PCal calibration have shown that the relative humidity variations around the PCal benches are correlated with the photodiodes calibration variations. 

\begin{figure}[h!] 
 \begin{center}
 	\captionsetup{justification=justified}
    \subfigure[NE PD1 O3a]{
	   \includegraphics[trim={0cm 0cm 0cm 0cm},clip,scale=0.43]{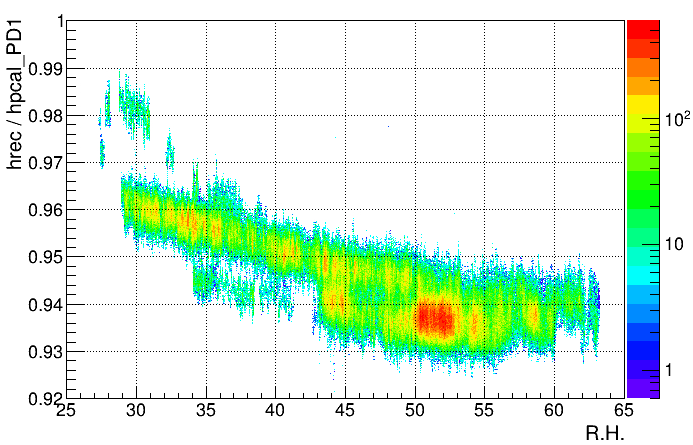}  
	   \label{fig:NEPD1RHO3a} }
    \subfigure[NE PD2 O3a]{
       \includegraphics[trim={0cm 0 0cm 0cm},clip,scale = 0.43]{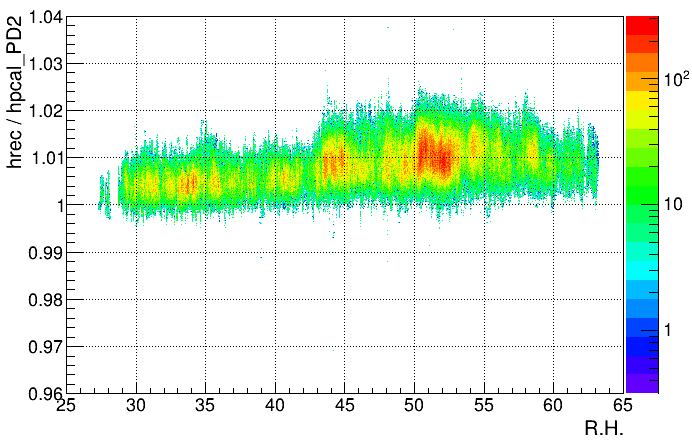}
       \label{fig:NEPD2RHO3a}  }
           \subfigure[NE PD1 O3b]{
	   \includegraphics[trim={0cm 0cm 0cm 0cm},clip,scale=0.43]{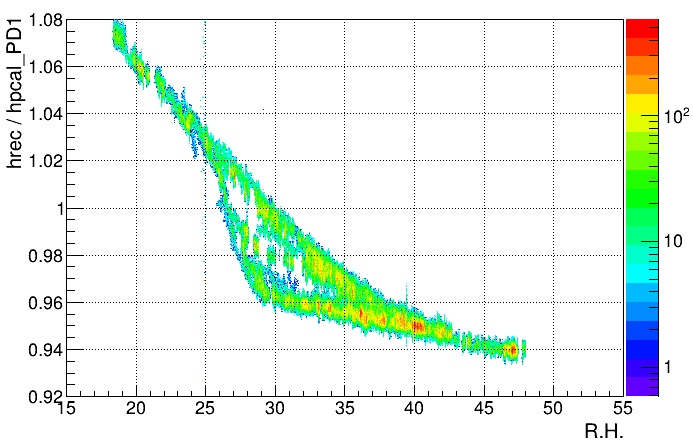}  
	   \label{fig:NEPD1RHO3b} }
    \subfigure[NE PD2 O3b]{
       \includegraphics[trim={0cm 0cm 0cm 0cm},clip,scale=0.43]{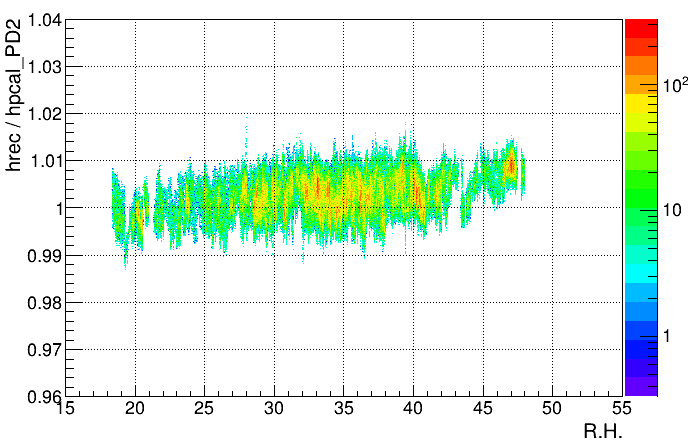}
       \label{fig:NEPD2RHO3b}  }
	\caption{Distribution of $h_{rec}/h_{pcal}$ for NE PCal photodiodes as a function of relative humidity (R.H.) surrounding NE PCal during O3a and O3b.}
	\label{fig:NERHvariations}
	\end{center}
\end{figure}

\begin{table}[h!]
\centering
\begin{tabular}{|c|c|c|c|c|}
\hline 
\multirow{2}{*}{Parameter} & \multicolumn{2}{c|}{$1\sigma$ uncertainty O3a}  & \multicolumn{2}{c|}{$1\sigma$ uncertainty O3b}\\ 
 & NE & WE & NE & WE \\ 
 \hline
Responsivity (temperature) & $\pm 0.1\%$ & $\pm 0.1\%$ & $\pm 0.1\%$ & $\pm 0.1\%$ \\ 
Other sources & $\pm 0.5\%$ & $\pm 0.5\%$ & $\pm 0.6\%$ & $\pm 1.2\%$ \\
\hline  
Total & $0.51\%$ & $0.51\%$ & $\pm 0.61\%$ & $\pm 1.2\%$\\
\hline
\end{tabular}
\caption{\label{tab:stabilityunc} Uncertainty on the stability in time of the PCal-induced end mirror displacement due to temperature-dependent photodiode response and other sources of uncertainty during O3.}
\end{table}
\newpage
\paragraph{Relative humidity variations affecting the PCal calibration}
~~\\\\
We have observed that the variations of humidity inside the NE PCal bench was correlated with the NE PD1 signal during O3a and O3b as shown in figure \ref{fig:NERHvariations}. The correlation varies in time and the behaviors during O3a and O3b are different. During O3a, at least two bands of correlation can be seen whereas during O3b a phenomenon of hysteresis has been observed. We thus suspect that the change in relative humidity is the main source of the amplitude variations of the NE PD1 response up to $\sim13\%$. However the variations of NE PD2 signal with respect to humidity variations are consistent with the $0.5\%$ and $0.6\%$ uncertainty estimated for the stability in time of its calibration over O3a and O3b.

\begin{figure}[h!] 
 \begin{center}
 	\captionsetup{justification=justified}
    \subfigure[WE PD1 O3a]{
	   \includegraphics[trim={0cm 0cm 0cm 0cm},clip,scale=0.43]{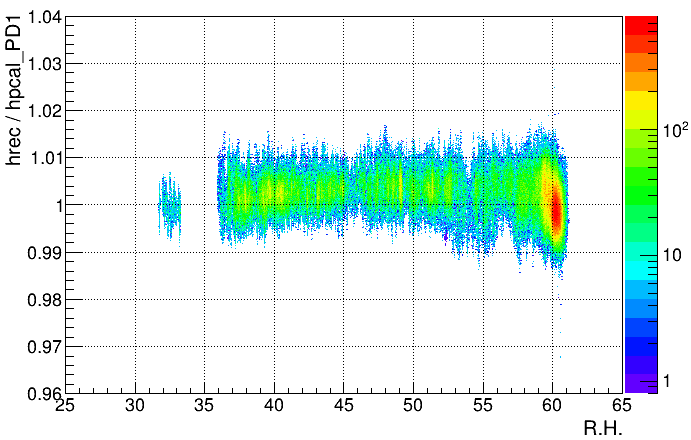}  
	   \label{fig:WEPD1RHO3a} }
    \subfigure[WE PD2 O3a]{
       \includegraphics[trim={0cm 0 0cm 0cm},clip,scale = 0.43]{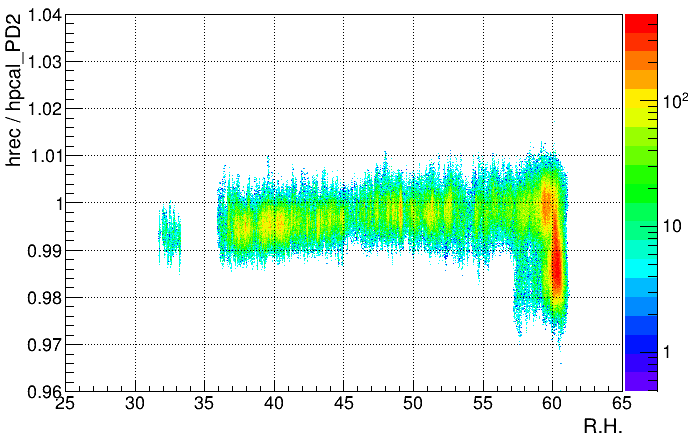}
       \label{fig:WEPD2RHO3a}  }
           \subfigure[WE PD1 O3b]{
	   \includegraphics[trim={0cm 0cm 0cm 0cm},clip,scale=0.43]{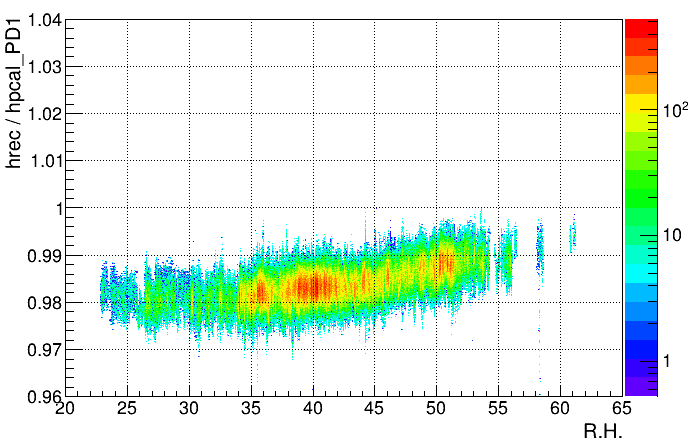}  
	   \label{fig:WEPD1RHO3b} }
    \subfigure[WE PD2 O3b]{
       \includegraphics[trim={0cm 0cm 0cm 0cm},clip,scale=0.43]{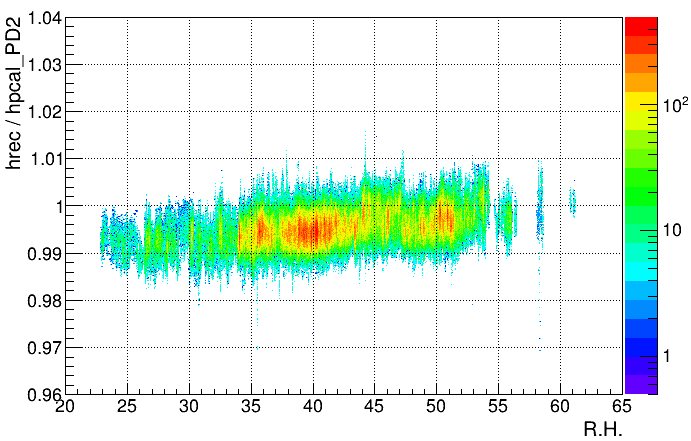}
       \label{fig:WEPD2RHO3b}  }
	\caption{Distribution of $h_{rec}/h_{pcal}$ for WE PCal photodiodes as a function of the relative humidity (R.H.) surrounding WE PCal during O3a and O3b.}
	\label{fig:WERHvariations}
	\end{center}
\end{figure}

Similar investigations have been performed on WE PCal photodiodes and are shown in figure \ref{fig:WERHvariations}. During O3a, WE PD1 signal variations due to humidity changes are consistent with the $0.5\%$ uncertainty given for its calibration stability in time. It is also noticeable that the main part of the $1.5\%$ variations of WE PD2 signal during O3a is not correlated with humidity changes and that humidity variations may count only for $0.5\%$ of the WE PD2 photodiode calibration variations. Regarding O3b, the variations of WE PD2 signal due to humidity changes were also within $\sim0.5\%$. 
 
After the recalibration of WE PD1 ($+1.3\%$) performed with a relative humidity of $60\%$ before O3b, the value of $h_{rec}/h_{pcal}$ differed by $0.5\%$ from the one of O3a at the same relative humidity (see figures \ref{fig:WEPD1RHO3a} and \ref{fig:WEPD1RHO3b}). This indicates that the calibration measurements of WE PD1 performed before O3a and a few days before O3b are coherent and compatible with the uncertainty of $0.8\%$ on VIS positioning on the optical bench. However, during O3b, WE PD1 calibration started to drift by $\sim1.1\%$ due to humidity variations. Adding quadratically the uncertainty of $0.5\%$ around the mean value, the uncertainty on WE PCal calibration stability for O3b has been increased to $1.2\%$.

\subsection{Error budget}
\begin{figure}[!h]
	\center
	\captionsetup{justification=justified}
	\includegraphics[trim={0cm 0 0 1cm},clip,scale=0.8]{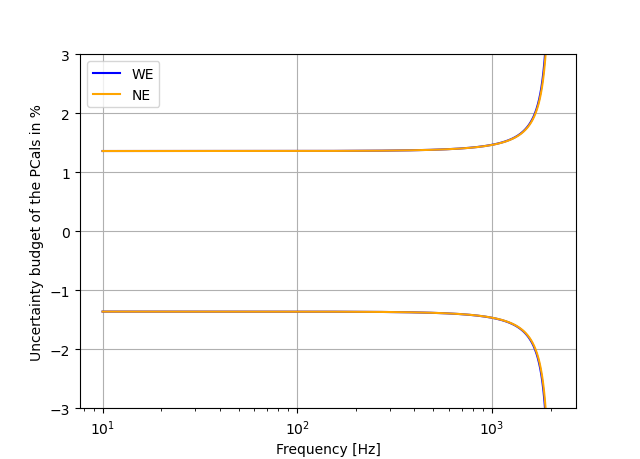} 
	\caption{Frequency dependent total uncertainty on the NE and WE mirrors displacement induced by the photon calibrators for O3a.}
	\label{fig:uncbudgetpcal}
\end{figure}

Below the notch frequency, the total uncertainty on the NE (resp. WE) photon calibrators has been estimated to be $1.36\%$ (resp. $1.36\%$) for O3a and $1.40\%$ (resp. $1.74\%$) for O3b. The detailed contributions to this uncertainty have been given in the previous sections and are summarized in table \ref{uncertainty}. For frequencies around the notch frequency, the interaction with the internal deformations of the optic strongly reduce the effective displacement sensed by the interferometer and the PCal uncertainty diverges. Figure \ref{fig:uncbudgetpcal} shows the uncertainty budget of both PCals for O3a from $10~$Hz to $2~$kHz. 

\begin{table}[!h]
\centering
\captionsetup{justification=justified}
\begin{tabular}{|c|c|c|c|c|}
\hline 
\multirow{2}{*}{Parameter} & \multicolumn{2}{c|}{$1\sigma$ uncertainty O3a}  & \multicolumn{2}{c|}{$1\sigma$ uncertainty O3b}\\ 
 & NE & WE & NE & WE \\ 
\hline 
Reflected laser power ($P$) & $1.24\%$ & $1.24\%$ & $1.24\%$ & $1.24\%$ \\ 
Geometrical parameters & $0.20\%$ & $0.20\%$ & $0.20\%$ & $0.20\%$  \\ 
Calibration stability (O3) & $0.51\%$ & $0.51\%$ & $0.61\%$ & $1.2\%$\\
\hline 
Total & $1.36\%$ & $1.36\%$ & $1.40\%$ & $1.74\%$ \\
\hline
\end{tabular}
\caption{\label{uncertainty} Uncertainty budget of the photon calibrators below $1~$kHz.}
\end{table}

In addition, a PCal timing uncertainty must be taken into account. Indeed, the signal measured by the photodiode to estimate the laser power reflected onto the end mirror is processed by a sensing path divided into an analog part and a digital part inducing a delay. Thanks to the measurements done with the 1PPS signal sent with a flashing light diode (see section \ref{subsec:timing}), this delay has been estimated to be $110\pm 3~\upmu$s, consistent with the expected value and where the main contribution comes from the digital part of the sensing path. This delay has been corrected to get the absolute timing of the PCal-induced end mirror motion during O3.

\section{Conclusion}
\label{sec:conclusion}

The photon calibrators developed for several years in Virgo and used in a preliminary version during O2, have been improved and used for the first time as a calibration reference during the observation run O3. This allowed to put the relative calibration of the gravitational wave detectors network on the same \textit{absolute} calibration reference: the Gold Standard. The first work of intercalibration between Virgo and LIGO PCals allowed us to correct for a discrepancy of $3.92\%$ on the measured laser power between the detectors. A Working Standard for Virgo, similar to the Working Standards used in LIGO and KAGRA, has also been mounted to check the stability in time of the Virgo Integrating Sphere calibration.

On the Advanced Virgo PCal, the laser power has been digitally controlled in order to keep its broadband noise contribution more than 10 times below the sensitivity of the Advanced Virgo interferometer. In addition, the systematic uncertainty on the PCal-induced end mirror motion of the interferometer has been estimated to be $1.36\%$ for both PCal during O3a and $1.40\%$ (resp. $1.74\%$) on NE PCal (resp. WE PCal) during O3b.
There is a common main contribution to the final systematic uncertainty on both PCals: the estimation of the PCal laser power reflected by the end mirror at the level of $1.24\%$. It will have to be reduced in the future and a first step will be to characterize the calibration of the Virgo Integrating Sphere as a function of the laser beam size. Then, it will be also convenient to use a larger beam on the \textit{injection bench} with a waist located on the interferometer's end mirror so that the beam on the \textit{injection} and \textit{reflection} benches has approximately the same size. Moreover, the Si photodiodes used to estimate the laser power reflected by the end mirror should be replaced by InGaAs photodiodes whose responsivities would have a smaller dependency on temperature variations on the PCal benches. During O3a, the stability of the photodiodes calibration has been estimated to $0.5\%$ on WE and NE PCals but bigger variations, correlated to humidity variations, have been seen on one of the NE photodiode on the \textit{injection} bench. During O3b, NE PD2 calibration stability was updated to $0.6\%$ with variations due to change in the relative humidity similar to the ones seen during O3a. However WE PD1 calibration started to experience changes during O3b up to $1.1\%$. This result urged us to increase the systematic uncertainty on the stability of the WE PCal calibration up to $1.2\%$. The cause of the calibration variations correlated to relative humidity changes has not yet been well understood but investigations are on-going. This phenomenon will have to be addressed in the future to improve the stability in time of the photodiodes calibration.

Not only the amplitude of the PCal-induced displacement needs to be calibrated but also its timing which is a critical feature for coincident data analysis between the different gravitational wave detectors. Using the LED flashing a 1PPS signal on the PCals photodiodes the \textit{absolute} timing of the PCal-induced motion has been measured to $110\pm3~\upmu$s and taken into account in the reconstruction of the gravitational wave signal.

The future improvements foreseen for the PCal stability or for the PCal laser reflected power accuracy will help reducing the uncertainty of the online $h(t)$ provided to data analysis. For the next observing runs, we can expect to reach and keep below $1\%$ the uncertainty for the PCal-induced displacement of the end mirrors.

\section*{Acknowledgements}
The authors gratefully acknowledge the University of Grenoble Alps Excellence Initiative (IDEX) and the United States National Science Foundation (NSF) for funding the D. Estevez travel grant to LIGO Hanford Observatory (LHO). The authors gratefully acknowledge R. L. Savage and Y. Lecoeuche for their contribution to the lab measurements performed at LHO for the intercalibration of the LIGO and Virgo PCals. The authors also gratefully acknowledge the  Italian Istituto  Nazionale  di  Fisica  Nucleare (INFN), the  French  Centre  National  de  la  Recherche  Scientifique  (CNRS)  and  the Foundation  for  Fundamental  Research  on  Matter  supported  by  the  Netherlands Organisation  for  Scientific  Research,  for  the  construction  and  operation  of  the  Virgo detector and the creation and support of the EGO consortium.

Part of the data used in this paper will become public in GWOSC (\url{https://www.gw-openscience.org/}): it is the case of the strain $h(t)$ channel ; the other channels will not be made public since they require a detailed knowledge of the detector to be correctly interpreted and used
(mainly calibration injection signals, PCal control channels, environment and monitoring channels).
We thank the Virgo collaboration for letting us using the data from the Virgo detector used in the paper.

\section*{References}

\bibliographystyle{iopart-num}
\bibliography{references}

\providecommand{\newblock}{}
\begin{thebibliography}{10}
\expandafter\ifx\csname url\endcsname\relax
  \def\url#1{{\tt #1}}\fi
\expandafter\ifx\csname urlprefix\endcsname\relax\def\urlprefix{URL }\fi
\providecommand{\eprint}[2][]{\url{#2}}

\bibitem{GWTC}
Abbott B~P {\em et~al.\/} (LIGO Scientific Collaboration and Virgo
  Collaboration) 2019 {\em Phys. Rev. X\/} {\bf 9} 031040
  \urlprefix\url{https://doi.org/10.1103/PhysRevX.9.031040}

\bibitem{H0}
Abbott B~P {\em et~al.\/} (LIGO Scientific Collaboration and Virgo
  Collaboration) 2017 {\em Nature\/} {\bf 551} 85--88
  \urlprefix\url{https://doi.org/10.1038/nature24471}

\bibitem{TGR}
Abbott B~P {\em et~al.\/} (LIGO Scientific Collaboration and Virgo
  Collaboration) 2019 {\em Phys. Rev. D\/} {\bf 100} 104036
  \urlprefix\url{https://doi.org/10.1103/PhysRevD.100.104036}

\bibitem{O2virgocalib}
Acernese F {\em et~al.\/} (Virgo collaboration) 2018 {\em Class. Quant.
  Grav.\/} {\bf 35} 205004
  \urlprefix\url{https://doi.org/10.1088%2F1361-6382%2Faadf1a}

\bibitem{O3calibligo}
Sun L {\em et~al.\/} 2020 {\em Class. Quant. Grav.\/} {\bf 37} 225008
  \urlprefix\url{https://doi.org/10.1088%2F1361-6382%2Fabb14e}

\bibitem{tripledetect}
Abbott B~P {\em et~al.\/} (LIGO Scientific Collaboration and Virgo
  Collaboration) 2017 {\em Phys. Rev. Lett.\/} {\bf 119}(14) 141101
  \urlprefix\url{https://link.aps.org/doi/10.1103/PhysRevLett.119.141101}

\bibitem{binaryneutron}
Abbott B~P {\em et~al.\/} (LIGO Scientific Collaboration and Virgo
  Collaboration) 2017 {\em Phys. Rev. Lett.\/} {\bf 119}(16) 161101
  \urlprefix\url{https://link.aps.org/doi/10.1103/PhysRevLett.119.161101}

\bibitem{multimessenger}
Abbott B~P {\em et~al.\/} 2017 {\em The Astrophysical Journal\/} {\bf 848} L12
  \urlprefix\url{https://doi.org/10.3847%2F2041-8213%2Faa91c9}

\bibitem{these_germain}
Germain V 2017 {\em De l'etalonnage d'Advanced Virgo a la recherche d'ondes
  gravitationnelles emises par des coalescences binaires compactes\/} Ph.D.
  thesis Universite Grenoble-Alpes

\bibitem{these_estevez}
Estevez D 2020 {\em Upgrade of Advanced Virgo photon calibrators and first
  intercalibration of Virgo and LIGO detectors for the observing run O3\/}
  Ph.D. thesis Universite Savoie Mont-Blanc

\bibitem{O3ligopcal}
Bhattacharjee D {\em et~al.\/} 2020  (\textit{Preprint} \eprint{2006.00130})
  \urlprefix\url{https://arxiv.org/abs/2006.00130}

\bibitem{aligopcal}
Karki S {\em et~al.\/} 2016 {\em Review of Scientific Instruments\/} {\bf 87}
  114503 ISSN 1089-7623 \urlprefix\url{http://dx.doi.org/10.1063/1.4967303}

\bibitem{Hildpcal}
Hild S {\em et~al.\/} 2007 {\em Classical and Quantum Gravity\/} {\bf 24}
  5681–5688 ISSN 1361-6382
  \urlprefix\url{http://dx.doi.org/10.1088/0264-9381/24/22/025}

\bibitem{GSnist}
{LIGO Pcal Group} 2018  \urlprefix\url{https://dcc.ligo.org/LIGO-T1900097}

\bibitem{typeunc}
Taylor B~N and Kuyatt C~E 1994 {\em NIST technical document\/} {\bf 1297}
  \urlprefix\url{https://emtoolbox.nist.gov/Publications/NISTTechnicalNote1297s.pdf}

\end{thebibliography}

\end{document}